\documentclass[twocolumn,floats,showpacs,superscriptaddress,pre]{revtex4}
\usepackage{amssymb}
\usepackage{graphicx}
\usepackage{dcolumn}
\usepackage{amsmath}
\usepackage{bm}
\usepackage{epsfig}

\newcommand{\be}{\begin{equation}}
\newcommand{\ee}{\end{equation}}

\newcommand{\media}[1]{\langle #1 \rangle}

\begin{document}

\title{Exact results and new insights
for models defined over small-world networks. \\
First and second order phase transitions. I: General result}
\author{M. Ostilli }
\affiliation{Departamento de F{\'\i}sica da Universidade de Aveiro, 3810-193 Aveiro,
Portugal}
\affiliation{Center for Statistical Mechanics and Complexity, 
INFM-CNR SMC, Unit\`a di Roma 1, Roma, 00185, Italy.}
\author{J. F. F. Mendes}
\affiliation{Departamento de F{\'\i}sica da Universidade de Aveiro, 3810-193 Aveiro,
Portugal}

\begin{abstract}
We present, as a very general method, 
an effective field theory to analyze
models defined over small-world networks.
Even if the exactness of the method is limited
to the paramagnetic regions and to some special limits, it gives the exact critical
behavior and the exact critical surfaces and percolation thresholds, 
and provide a clear and immediate (also in terms of calculation)
insight of the physics. 
The underlying structure of the non random part of the model, \textit{i.e.},
the set of spins staying in a given lattice $\mathcal{L}_0$ of dimension $d_0$
and interacting through a fixed coupling $J_0$, is exactly
taken into account. When $J_0\geq 0$, the small-world effect gives rise
to the known fact that a second order phase transition takes place,
independently of the dimension $d_0$ and of the added random connectivity $c$.
However, when $J_0<0$, a completely different scenario emerges 
where, besides a spin glass transition, 
multiple first- and second-order phase transitions may take place.
\end{abstract}

\maketitle

\email{ostilli@roma1.infn.it}

\section{Introduction} \label{intro}
Since the very beginning of the pioneeristic 
work by Watts and Strogatz \cite{Watts}, the interest toward small-world
networks - an interplay between random and regular networks -
has been growing ``exponentially''.
Mainly, there are two reasons that have caused such a ``diffusion''.

The first reason is due to the topological properties of the small-world
network. In synthesis, if $N$ is the size of the system, 
for any finite probability $p$ of rewiring, or for any finite added random
connectivity $c$       
(the two situations 
correspond to two slightly different procedures for building a small-world
network) one has: 
a ``short-distance behavior'', implying that 
the shortest distance between two arbitrarily chosen
sites grows as $l(N)\sim \log(N)$, as in random networks, 
and a large clustering coefficient, $C(N)\sim\mathop{O}(1)$, as in regular lattices.
The interplay between these two features makes small-world networks
representative of many realistic situations ranging from social networks, 
communications networks, chemical reactions networks, protein networks, 
neuronal networks, etc.

The second reason is due the fact that, 
in models defined over small-world networks, despite 
the presence of an underlying finite dimensional structure - a lattice
$\mathcal{L}_0$ of dimension $d_0<\infty$ - the existence of the
short-cut bonds makes such models mean-field-like and - hopefully - exactly solvable. 
However, even if such a claim sounds intuitively
correct, the complexity of these models turns out to be in general quite high
and, compared to numerical works, analytical results, and especially exact results, 
on small-world networks are still few \cite{Barrat}-\cite{Bolle} 
(for percolation and synchronization problem we cite in particular
\cite{Newman} and \cite{Hastings3}).

In particular, for $d_0>1$ an analytical approach
seems to be impossible, even though, again, a mean-field critical behavior
is expected and has been already confirmed by Monte Carlo (MC) simulations \cite{Herrero}.
A natural question is, are we able to prove analytically such an assertion?
If for example $d_0=2$, does the mean-field critical behavior hold for any situation?
Yet, does the correlation length diverge at the critical temperature?

Furthermore, even if for $d_0=1$ an exact analytical treatment has been
performed at the level of replica symmetry (RS) \cite{Niko} and one step replica symmetry
breaking (1RSB) \cite{Niko2}, the calculations involved are quite long
and the solutions of the coupled equations for evaluating the order parameters
require a certain numerical work becoming rapidly hard in the 1RSB case.
In any case, even if
these methods 
are able to give in principle exact results at any temperature,  
they are not in general suitable for giving a clear simple and immediate
physical picture of the model, 
even - possibly - within some approximation. The main problem in fact
stays in the presence of the short-loops: as soon as $d_0>1$ these loops
cannot be neglected and the ``traditional'' cavity and replica methods seem hardly applicable.
In particular, what happens, for example, if we set $J_0$ negative?
Shall we still expect a second order phase transition? 
And what about the phase diagram?

In this paper, we present a general method to study random Ising models defined
on small-world graphs built up by adding a random connectivity $c$ over
an underlying arbitrary lattice $\mathcal{L}_0$ having dimension $d_0$.
We shall then show that this method provides - in a very simple and physically
appealing way - the answers to the above questions and many others.
 
As an effective field theory, roughly speaking, the method generalizes 
the mean-field equation $m=\tanh(\beta J m)$ to take into account
the presence of the short-range couplings $J_0$ besides the long-range ones $J$.
As we will show, the magnetization $m$ of the model defined over the
small-world network, shortly the \textit{random model}, behaves as the magnetization
$m_0$ of the model defined over $\mathcal{L}_0$, shortly the \textit{unperturbed model},
but immersed in an effective field to be determined self-consistently.
Even if the exactness of this method is limited
to the paramagnetic regions (P), it gives the exact critical
behavior and the exact critical surfaces, 
and provide simple qualitative good estimations of
the correlation functions in the ferromagnetic (F) and spin glass regions (SG).
Furthermore, in unfrustrated systems, the method becomes exact at any temperature
in the two limits $c\to 0^+$ and $c\to\infty$.

The consequences of such a general result are remarkable
from both the theoretical and the practical point of view.
Once the explicit form of the magnetization of the unperturbed model, 
$m_0=m_0(\beta J_0,\beta h)$, 
as a function of the couplings $J_0$ and of the external field $h$ is known,
analytically or numerically,
we get an approximation to the full solution of the random model analytically or numerically,
respectively, becoming exact in the P region.
If we do not have $m_0=m_0(\beta J_0,\beta h)$ but we know at least 
some of its properties, 
we can still use these properties to derive certain exact relations and the
critical behavior of the random model.

In this paper (part I), after discussing the self-consistent equations,
we will mainly focus on applying them to study the critical surfaces
and the critical behavior in general,
whereas in a forthcoming paper (part II), after showing some examples, we will
apply the method to study several models of interest which can be solved 
analytically (and very easily) as for them we know $m_0(\beta J_0,\beta h)$. 

We stress that, the critical surfaces, as well as the correlation functions 
in the P region provided by our method, are exact and
- quite interestingly -
not based on any special ansatz as the replica-symmetry and the tree-like
ansatz.
We prove in particular that: for $J_0\geq 0$, independently of the added
random connectivity $c$ and of the underlying dimension $d_0$,
we always have a second order phase transition with the classical mean-field critical
indices but with a finite correlation length if calculated along the distance
defined by the underlying lattice $\mathcal{L}_0$; whereas for $J_0<0$,
we show that, as soon as $c$ is sufficiently large, there exist at least two critical
temperatures which, depending on the behavior of $\chi_0(\beta J_0,\beta h)$
 - the susceptibility of the unperturbed system - 
correspond to first or second order phase transitions. 

The paper is organized as follows.
In Sec. II we introduce the class of small-world networks over which
we define the random Ising models, stressing some important 
difference concerning the definition of the correlation functions 
with respect to the definition of the correlation functions one usually 
considers in ``ordinary'' random models. 
In Sec. III we present our method: in Sec. IIIA we provide the self-consistent
equations and their relations with physical correlation functions,
in Sec. IIIB we analyze the stability of the solutions of the self consistent
equations and the critical surface and behavior of the system. We separate 
the Sec. IIIB in the sub-cases $J_0\geq 0$ and $J_0<0$. 
In Sec. IIIC we discuss the limits of the method. 
In Sec. IIID we study the stability between the F and the SG phases
and the phase diagram.
Finally, in Sec. IIIE we mention how to generalize the method 
to cases in which there are more different short-range couplings $J_0$, 
and how to generalize the method to analyze possible disordered antiferromagnetism.
The successive Secs. IV, V and VI are devoted to the derivation of the method.
The starting point of the proof is given in Sec. IV and is 
based on a general mapping between a random model and a non random one
\cite{MOI}-\cite{MOIII} suitably adapted to our case.
The self-consistent equations are then easily derived in Sec. V. 
Note that, apart from the equations concerning the stability between the P-F and
the P-SG transitions, which are derived in Sec. VI, the derivations of
the equations provided in the Sec. IIIB are mostly left to the reader, since
they can be easily obtained by standard arguments of statistical mechanics 
using the Landau free energy
$\psi(m)$ that we provide and that is derived in Sec. V too.
Finally, in Sec. VII we draw some conclusions.
In Appendix A we generalize the method to inhomogeneous external fields
to make clear the subtle behavior of the correlation functions 
in small-world models.
%
%
%

\section{Random Ising models on small-world networks}
\label{models}
The family of models we shall consider are random Ising models
constructed by super-imposing random graphs with finite average connectivity $c$
onto some given lattice $\mathcal{L}_0$ whose set of bonds $(i,j)$ 
and dimension will be indicated by $\Gamma_0$ and $d_0$, respectively.
Given an Ising model (\textit{the unperturbed model}) 
of $N$ spins coupled over $\mathcal{L}_0$
through a coupling $J_0$  
with Hamiltonian
\begin{eqnarray}
H_0\equiv -J_{0}\sum_{(i,j)\in \Gamma_0}\sigma_{i}\sigma_{j}-h\sum_i \sigma_i
\label{H0},
\end{eqnarray}
and given an ensemble $\mathcal{C}$ 
of unconstrained random graphs $\bm{c}$, $\bm{c}\in\mathcal{C}$,
whose bonds are determined by the adjacency matrix elements $c_{i,j}=0,1$,
we define the corresponding small-world model  
as described by the following Hamiltonian
\begin{eqnarray}
\label{H}
H_{\bm{c},\bm{J}}\equiv H_0-\sum_{i<j} c_{ij}{J}_{ij}\sigma_{i}\sigma_{j},
\end{eqnarray}
the free energy $F$ and the averages $\overline{\media{\mathcal{O}}^l}$
being defined in the usual (quenched) way as ($\beta=1/T$)
\begin{eqnarray}
\label{logZ}
-\beta F\equiv \sum_{\bm{c}\in\mathcal{C}} 
P(\bm{c})\int d\mathcal{P}\left(\{{J}_{i,j}\}\right)
\log\left(Z_{\bm{c},\bm{J}}\right)
\end{eqnarray} 
and 
\begin{eqnarray}
\label{O}
\overline{\media{\mathcal{O}}^l}\equiv 
\sum_{\bm{c}\in\mathcal{C}} P(\bm{c}) \int d\mathcal{P}\left(\{{J}_{i,j}\}\right)
\media{\mathcal{O}}_{\bm{c},\bm{J}}^l, \quad l=1,2
\end{eqnarray} 

where $Z_{\bm{c},\bm{J}}$ 
is the partition function of the quenched system

\begin{eqnarray}
\label{Z}
Z_{\bm{c},\bm{J}}= \sum_{\{\sigma_{i}\}}
e^{-\beta H_{\bm{c},\bm{J}}\left(\{\sigma_i\}\right)}, 
\end{eqnarray} 

$\media{\mathcal{O}}_{\bm{c},\bm{J}}$ the Boltzmann-average 
of the quenched system ($\media{\mathcal{O}}$ depends on the
given realization of the ${J}$'s and of $\bm{c}$:
$\media{\mathcal{O}}=\media{\mathcal{O}}_{\bm{c};\bm{J}}$;
for shortness we later will omit to write these dependencies)

\begin{eqnarray}
\label{OO}
\media{\mathcal{O}}_{\bm{c},\bm{J}}\equiv \frac{\sum_{\{\sigma_i\}}\mathcal{O}e^{-\beta 
H_{\bm{c},\bm{J}}\left(\{\sigma_i\}\right)}}{Z_{\bm{c},\bm{J}}}, 
\end{eqnarray} 

and $d\mathcal{P}\left(\{{J}_{i,j}\}\right)$ and $P(\bm{c})$ 
are two product measures 
given in terms of two normalized measures $d\mu(J_{i,j})\geq 0$ and $p(c_{i,j})\geq 0$, respectively: 
\begin{eqnarray}
\label{dP}
d\mathcal{P}\left(\{{J}_{i,j}\}\right)\equiv \prod_{(i,j),i<j} 
d\mu\left( {J}_{i,j} \right),
\quad \int d\mu\left( {J}_{i,j} \right)=1,
\end{eqnarray}
\begin{eqnarray}
\label{Pg}
P(\bm{c})\equiv \prod_{(i,j),i<j} p(c_{i,j}),
\quad \sum_{c_{i,j}=0,1} p(c_{i,j})=1.
\end{eqnarray}

The variables
$c_{i,j}\in\{0,1\}$ specify whether a ``long-range'' bond between the sites
$i$ and $j$ is present ($c_{i,j}=1$) or absent ($c_{i,j}=0$), whereas
the $J_{i,j}$'s are the random variables of the given bond $(i,j)$.
For the $J_{i,j}$'s we will not assume any particular distribution, 
while, to be specific, for the $c_{i,j}$'s we shall consider the following 
distribution
\begin{eqnarray}
\label{PP}
 p(c_{ij})=
\frac{c}{N}\delta_{c_{ij},1}+\left(1-\frac{c}{N}\right)\delta_{c_{ij},0}.
\end{eqnarray}
This choice leads in the thermodynamic limit $N\to\infty$ to a
number of long range connections per site distributed according
to a Poisson law with mean $c>0$ (so that in average there are in total $cN/2$ bonds). 
Note however that the main results we report
in the next section are easily generalizable to any case in which
Eq. (\ref{Pg}) holds, or holds only in the thermodynamic limit due
a sufficiently small number of constrains among the matrix elements
$c_{i,j}$. 

The quantities of major interest are the averages, and the quadratic
averages, of the correlation functions 
which for shortness will be indicated by ${{C}}^{(\mathrm{1})}$ and
${{C}}^{(\mathrm{2})}$. For example, the following are non connected correlation functions
of order $k$:
\begin{eqnarray}
\label{CF}
{{C}}^{(\mathrm{1})}&=&\overline{\media {\sigma_{i_1}\ldots \sigma_{i_k}} }, \\
\label{CG}
{{C}}^{(\mathrm{2})}&=&\overline{\media {\sigma_{i_1}\ldots \sigma_{i_k}}^2 },
\end{eqnarray} 
where $k\geq 1$ and the indices $i_1,\ldots,i_k$ are supposed all different. 
For shortness we will keep on to use the symbols ${{C}}^{(\mathrm{1})}$ 
and ${{C}}^{(\mathrm{2})}$
also for the connected correlation function since, as we shall see in the next
section, they obey to the same rules of transformations.
We point out that the set of indices $i_1,\ldots,i_k$ is
fixed along the process of the two averages. This implies in particular
that, if we consider the spin with index $i$ and the spin with index $j$, 
their distance remains undefined, or more precisely, 
the only meaningful distance between $i$ and $j$, is the distance  
defined over $\mathcal{L}_0$, \textit{i.e.}, the Euclidean distance
between $i$ and $j$, which we will indicate as $||i-j||_{_{_0}}$.
 
Therefore, throughout this paper, it must be kept in mind that, 
for example, ${{C}}^{(\mathrm{1})}(||i-j||_{_{_0}})=
\overline{\media{\sigma_{i}\sigma_{j}} }$ 
is very different from the correlation function ${{G}}^{(\mathrm{1})}(l)$ 
of two points at a fixed
distance $l$, $l$ being here the distance defined over both $\mathcal{L}_0$
and the random graph $\bm{c}$, \textit{i.e.}, the minimum number 
of bonds to join two points
among both the bonds of $\Gamma_0$ and the bonds of the random graph $\bm{c}$.
In fact, if, for $J_0=0$, one considers all the possible 
realizations of the Poisson graph, and then all the possible
distances $l$ between two given points $i$ and $j$, one has   
\begin{eqnarray}
\label{CF1}
{{C}}^{(\mathrm{1})}(||i-j||_{_{_0}})&=&
\overline{\media {\sigma_{i}\sigma_{j}} }-\overline{\media
  {\sigma_{i}}} \overline{\media {\sigma_{j}} }
\nonumber \\
&=&\sum_{l=1}^N P_N(l){{G}}^{(\mathrm{1})}(l)
\end{eqnarray} 
where here $P_N(l)$ is the probability that, in the system with $N$ spins, 
the shortest path between the vertices $i$ and $j$ has length $l$.
  
If we now use ${{G}}^{(\mathrm{1})}(l)$
$\sim (\tanh(\beta J))^l$ \cite{Corr} (in the P region holds the equality) and the fact that
the average of $l$ with respect to $P_N(l)$ 
is of the order $\log(N)$, we see that the two point connected
correlation function (\ref{CF1}) goes to 0 in the thermodynamic limit. 
Similarly, all the connected
correlation functions defined in this way are zero in this limit.
Note however, that this independence of the variables holds only if $J_0=0$.
This discussion will be more deeply analyzed along the proof by using another point of view,
based on mapping the random model to a suitable fully connected model.


\section{An effective field theory}
\subsection{The self-consistent equations}
Depending on the temperature T, and 
on the parameters of the probability distributions, $d\mu$ and $p$,
the random model may stably stay either in the P, in the F, or in the SG phase.
In our approach for the F and SG phases there are two natural order parameters
that will be indicated by $m^{(\mathrm{F})}$ and $m^{(\mathrm{SG})}$.
Similarly, for any correlation function, quadratic or not, there are
two natural quantities 
indicated by $C^{(\mathrm{F})}$ and $C^{(\mathrm{SG})}$, and that in turn
will be calculated in terms of $m^{(\mathrm{F})}$ and $m^{(\mathrm{SG})}$,
respectively. 
To avoid confusion, it should be kept in mind that
in our approach,
for any observable $\mathcal{O}$,
there are - in principle - always 
two solutions that we label as F and SG,
but, as we shall discuss in Sec. IIID, for any temperature,
only one of the two solutions is stable and useful 
in the thermodynamic limit.

In the following, we will use the label $\mathop{}_0$ 
to specify that we are referring
to the unperturbed model with Hamiltonian (\ref{H0}).
Let $m_0(\beta J_0,\beta h)$ be the stable magnetization 
of the unperturbed model with coupling $J_0$ and in the presence of a uniform
external 
field $h$ at inverse temperature $\beta$. 
Then, the order parameters 
$m^{(\Sigma)}$, $\Sigma$=F,SG, 
satisfy the following self-consistent decoupled equations
\begin{eqnarray}
\label{THEOa}
m^{(\Sigma)}=m_0(\beta J_0^{(\Sigma)},
\beta J^{(\Sigma)}m^{(\Sigma)}+\beta h),
\end{eqnarray} 
where the effective couplings $J^{(\mathrm{F})}$, $J^{(\mathrm{SG})}$,
$J_0^{(\mathrm{F})}$ and $J_0^{(\mathrm{SG})}$ are given by
\begin{eqnarray}
\label{THEOb}
\beta J^{(\mathrm{F})}= c\int d\mu(J_{i,j})\tanh(\beta J_{i,j}),
\end{eqnarray} 
\begin{eqnarray}
\label{THEOc}
\beta J^{(\mathrm{SG})}= c\int d\mu(J_{i,j})\tanh^2(\beta J_{i,j}),
\end{eqnarray} 
\begin{eqnarray}
\label{THEOd}
J_0^{(\mathrm{F})}= J_0,
\end{eqnarray} 
and
\begin{eqnarray}
\label{THEOe}
\beta J_0^{(\mathrm{SG})}= \tanh^{-1}(\tanh^2(\beta J_0)).
\end{eqnarray}
Note that $|J_0^{(\mathrm{F})}|>J_0^{(\mathrm{SG})}$.

For the correlation functions ${{C}}^{(\Sigma)}$, $\Sigma$=F,SG, for sufficiently 
large $N$ we have
\begin{eqnarray}
\label{THEOh}
{{C}}^{(\Sigma)}=
{{C}}_0(\beta J_0^{(\Sigma)},\beta J^{(\Sigma)} m^{(\Sigma)}+\beta h)
+\mathop{O}\left(\frac{1}{N}\right),
\end{eqnarray} 
where ${{C}}_0(\beta J_0,\beta h)$ is the correlation function of the unperturbed 
(non random) model.

Concerning the free energy density $f$ we have 
\begin{eqnarray}
\label{THEOl} 
&& \beta f^{(\Sigma)} = -\frac{c}{2}\int d\mu(J_{i,j})
\log\left[\cosh(\beta J_{i,j})\right]
\nonumber \\
&& - \lim_{N\to\infty}\frac{1}{N}\sum_{(i,j)\in\Gamma_0}\log
\left[\cosh(\beta J_0)\right]
-\log\left[2\cosh(\beta h)\right]
\nonumber \\
&& + \{\lim_{N\to\infty}\frac{1}{N}\sum_{(i,j)\in\Gamma_0}\log
\left[\cosh(\beta J_0^{(\Sigma)})\right] 
\nonumber \\
&& +\log\left[2\cosh(\beta h)\right]\}
\times \frac{1}{l} +\frac{1}{l}L^{(\Sigma)}(m^{(\Sigma)}),
\end{eqnarray}
where $l=1,2$ for $\Sigma$=F,SG, respectively, and  
\begin{eqnarray}
\label{THEOll}
L^{(\Sigma)}(m)\equiv 
\frac{\beta J^{(\Sigma)}\left(m\right)^2}{2}+
\beta f_0\left(\beta J_0^{(\Sigma)},\beta J^{(\Sigma)}m+\beta h\right),
\end{eqnarray} 
$f_0(\beta J_0,\beta h)$ being the free energy density in the thermodynamic
limit of the unperturbed model with coupling $J_0$ and in the presence
of an external field $h$, at inverse temperature $\beta$.

For given $\beta$, among all the possible solutions of Eqs. (\ref{THEOa}), in the thermodynamic
limit, for both $\Sigma$=F and $\Sigma$=SG, 
the true solution $\bar{m}^{(\Sigma)}$, or leading solution, 
is the one that minimizes $L^{(\Sigma)}$:
\begin{eqnarray}
\label{THEOlead}
L^{(\Sigma)}\left(\bar{m}^{(\Sigma)}\right)&=&
\min_{m\in [-1,1]} L^{(\Sigma)}\left(m\right).
\end{eqnarray} 

Finally, let $k$ be the order of a given correlation function
$C^{(\mathrm{1})}$ or $C^{(\mathrm{2})}$.
The averages and the quadratic averages over the disorder,
$C^{(\mathrm{1})}$ and $C^{(\mathrm{2})}$, are related to  
$C^{(\mathrm{F})}$ and $C^{(\mathrm{SG})}$, as follows
\begin{eqnarray}
\label{THEOa0}
C^{(\mathrm{1})}&=&C^{(\mathrm{F})}, \quad \mathrm{in~F}, \\
\label{THEOa01}
C^{(\mathrm{1})}&=& 0, \quad k ~ \mathrm{odd}, \quad \mathrm{in~SG}, \\
\label{THEOa02}
C^{(\mathrm{1})}&=&C^{(\mathrm{SG})}, \quad k ~ \mathrm{even}, \quad \mathrm{in~SG},
\end{eqnarray} 
and
\begin{eqnarray}
\label{THEOa03}
C^{(\mathrm{2})}&=&\left(C^{(\mathrm{F})}\right)^2, \quad \mathrm{in~F}, \\
\label{THEOa04}
C^{(\mathrm{2})}&=&\left(C^{(\mathrm{SG})}\right)^2, \quad \mathrm{in~SG}.
\end{eqnarray} 

From Eqs. (\ref{THEOa03}) and (\ref{THEOa04}) for $k=1$, we note that
the Edward-Anderson order parameter
$C^{(\mathrm{2})}=\overline{\media{\sigma}^2}=q_{EA}$ is 
equal to $(C^{(\mathrm{SG})})^2=(m^{(\mathrm{SG})})^2$ only in the SG phase, whereas 
in the F phase we have $q_{EA}=(m^{(\mathrm{F})})^2$.
Therefore, since $m^{(\mathrm{SG})}\neq m^{(\mathrm{F})}$,
$m^{(\mathrm{SG})}$ is not equal to $\sqrt{q_{EA}}$;
in our approach $m^{(\mathrm{SG})}$ represents a sort of a spin glass order parameter.

The localization and the reciprocal stability between the F and SG phases will be
discussed in Sec. IIID. Note however that, at least for lattices $\mathcal{L}_0$
having only loops of even length, the stable P region is always that 
corresponding to a P-F phase diagram, so that in the P region
the correlation functions must be calculated only
through Eqs. (\ref{THEOa0}) and (\ref{THEOa03}).

As an immediate consequence of Eq. (\ref{THEOa}) we get the susceptibility 
$\tilde{\chi}^{(\Sigma)}$ of the random model:
\begin{eqnarray}
\label{THEOchie}
\tilde{\chi}^{(\Sigma)}= 
\frac{\tilde{\chi}_0\left(\beta J_0^{(\Sigma)},\beta J^{(\Sigma)}m^{(\Sigma)}+\beta h\right)}
{1-\beta J^{(\Sigma)}\tilde{\chi}_0
\left(\beta J_0^{(\Sigma)},\beta J^{(\Sigma)}m^{(\Sigma)} +\beta h\right)},
\end{eqnarray}
where $\tilde{\chi}_0$ stands for the susceptibility $\chi_0$ of the
unperturbed model divided by $\beta$ (we will adopt throughout this dimensionless 
definition of the susceptibility):
\begin{eqnarray}
\label{THEOchiedef}
\tilde{\chi}_0 \left(\beta J_0,\beta h\right)\equiv 
\frac{\partial m_0\left(\beta J_0,\beta h\right)}{\partial (\beta h)}
=\frac{1}{\beta}\frac{\partial m_0\left(\beta J_0,\beta h\right)}{\partial h},
\end{eqnarray}
and similarly for the random model.

For the case $\Sigma=$F without disorder ($d\mu(J')=\delta(J'-J)dJ'$), 
Eq. (\ref{THEOchie}) was already derived in \cite{Hastings2} by series expansion
techniques at zero field ($h=0$) in the P region (where $m=0$). 

Another remarkable consequence of our theory comes from Eq. (\ref{THEOh}).
We see in fact that in the thermodynamic limit any correlation function
of the random model fits with the correlation function of the unperturbed model
but immersed in an effective field that is exactly 
zero in the P region and zero external field ($h=0$).
In other words, in terms of correlation functions, 
in the P region, the random model and the unperturbed model are indistinguishable
(modulo the transformation $J_0\to J_0^{(\mathrm{SG})}$ for $\Sigma=$F).
Note however that this assertion holds only for a given correlation function 
calculated in the thermodynamic limit. In fact, the corrective $\mathop{O}(1/N)$ term
appearing in the rhs of Eq. (\ref{THEOh}) cannot be neglected when
we sum the correlation functions over all the sites $i\in\mathcal{L}_0$,
as to calculate the susceptibility; yet it is just
this corrective $\mathop{O}(1/N)$ 
term that gives rise to the singularities in the random model.

More precisely, for the two point connected correlation function: 
\begin{eqnarray}
\label{THEOC2}
\tilde{\chi}_{i,j}^{(\Sigma)}\equiv 
\overline{\media{\sigma_i\sigma_j}^l-\media{\sigma_i}^l\media{\sigma_j}^l},
\end{eqnarray}
where $l=1,2$ for $\Sigma=$ F, SG, respectively, if 
\begin{eqnarray}
\label{THEOC2a}
\tilde{\chi}_{0;i,j}\equiv 
\media{\sigma_i\sigma_j}_0-\media{\sigma_i}_0\media{\sigma_j}_0,
\end{eqnarray}
we have
\begin{eqnarray}
\label{THEOC2b}
\tilde{\chi}_{i,j}^{(\Sigma)}&=&
\tilde{\chi}_{0;i,j}(\beta J_0^{(\Sigma)},\beta J^{(\Sigma)} m^{(\Sigma)}+\beta h)+\frac{\beta J^{(\Sigma)}}{N}
\nonumber \\ &\times & \frac{\left[\tilde{\chi}_{o} 
(\beta J_0^{(\Sigma)},\beta J^{(\Sigma)}{{m}}^{(\Sigma)}+ \beta h)\right]^2}
{1-\beta J^{(\Sigma)} \tilde{\chi}_{o} 
(\beta J_0^{(\Sigma)},\beta J^{(\Sigma)}{{m}}^{(\Sigma)}+ \beta h)}.
\end{eqnarray}

Eq. (\ref{THEOC2b}) clarifies the structure of the correlation
functions in small-world models. In the rhs we have two terms: the former
is a distance-dependent short-range term whose finite correlation length, for $T\neq T_{c0}^{(\Sigma)}$ 
($T_{c0}^{(\Sigma)}$ being the critical temperature of the unperturbed model with coupling $J_0^{(\Sigma)}$),
makes it normalizable, the latter is instead a distance-independent long-range term which turns out
to be normalizable thanks to the $1/N$ factor. Once summed, both the terms give a finite
contribution to the susceptibility.
It is immediate to verify that by summing $\tilde{\chi}_{i,j}^{(\Sigma)}$ 
over all the indices $i,j\in\mathcal{L}_0$ and dividing by $N$ we get back - as it must be -
Eq. (\ref{THEOchie}). Eq. (\ref{THEOC2b}) will be derived in Appendix A where we generalize
the theory to a non homogeneous external field.

\subsection{Stability: critical surfaces and critical behavior}
Note that, for $\beta$ sufficiently small (see later), 
Eq. (\ref{THEOa}) has always the solution $m^{(\Sigma)}=0$, and furthermore,
if $m^{(\Sigma)}$ is a solution, $-m^{(\Sigma)}$ is a solution as well. 
From now on, if not explicitly said, 
we will refer only to the positive (possibly zero) solution,
the negative one being understood.
A solution $m^{(\Sigma)}$ of Eq. (\ref{THEOa}) is stable (but in general not unique) if
\begin{eqnarray}
\label{THEOO}
1- \beta J^{(\Sigma)}\tilde{\chi}_0 
\left(\beta J_0^{(\Sigma)},\beta J^{(\Sigma)}m^{(\Sigma)} + \beta h\right)>0.
\end{eqnarray}

For what follows, we need to rewrite
the non trivial part of the free energy density
$L^{(\Sigma)}(m)$ as
\begin{eqnarray}
\label{THEOL}
L^{(\Sigma)}(m)&=&
\beta f_0 \left(\beta J_0^{(\Sigma)},0\right) 
- m_0\left(\beta J_0^{(\Sigma)},0\right)\beta h 
\nonumber \\
&& + \psi^{(\Sigma)}\left(m\right),
\end{eqnarray}
where the introduced term $\psi^{(\Sigma)}$ plays the role of a Landau free energy
density and is responsible for the critical behavior of the system. 
Around $m=0$, up to terms $\mathop{O}(h^2)$ and $\mathop{O}(m^3 h)$,
$\psi^{(\Sigma)}(m)$ can be expanded as follows
\begin{eqnarray}
\label{THEOL1}
\psi^{(\Sigma)}\left(m\right)
&=& \frac{1}{2}a^{(\Sigma)} m^2 + 
\frac{1}{4} b^{(\Sigma)} m^4
+\frac{1}{6} c^{(\Sigma)} m^6
\nonumber \\ &&
 - m\beta \tilde{h}^{(\Sigma)}
\nonumber \\ &&
+\Delta\left(\beta f_0\right)(\beta J_0^{(\Sigma)},\beta J^{(\Sigma)}m),
\end{eqnarray}
where 
\begin{eqnarray}
\label{THEOL2}
a^{(\Sigma)}=
\left[1-\beta J^{(\Sigma)}\tilde{\chi}_0 
\left(\beta J_0^{(\Sigma)},0\right)\right]
\beta J^{(\Sigma)},
\end{eqnarray}
\begin{eqnarray}
\label{THEOL3}
b^{(\Sigma)}=- \frac{\partial^2}{\partial(\beta h)^2}
{\left. {\tilde{\chi}_0\left(\beta J_0^{(\Sigma)},\beta h\right)}
\right|_{_{\beta h =0} }
\frac{\left(\beta J^{(\Sigma)}\right)^4}{3!}},
\end{eqnarray}
\begin{eqnarray}
\label{THEOL4}
c^{(\Sigma)}=- \frac{\partial^4}{\partial(\beta h)^4}
{\left. {\tilde{\chi}_0\left(\beta J_0^{(\Sigma)},\beta h\right)}
\right|_{_{\beta h =0} }
\frac{\left(\beta J^{(\Sigma)}\right)^6}{5!}},
\end{eqnarray}
\begin{eqnarray}
\label{THEOL5}
\tilde{h}^{(\Sigma)}=m_0\left(\beta J_0^{(\Sigma)},0\right)J^{(\Sigma)}+
\tilde{\chi}_0\left(\beta J_0^{(\Sigma)},0\right)J^{(\Sigma)}h,
\end{eqnarray}
finally, the last term 
$\Delta \left(\beta f_0\right) \left(\beta J_0,\beta J^{(\Sigma)}m\right)$
is defined implicitly to render Eqs. (\ref{THEOL}) and (\ref{THEOL1}) exact, 
but terms $\mathop{O}(h^2)$ and $\mathop{O}(m^3 h)$, explicitly:
\begin{eqnarray}
\label{THEOL6}
&&\Delta \left(\beta f_0\right) \left(\beta J_0^{(\Sigma)},\beta J^{(\Sigma)}m\right)=
\nonumber \\
&-&\sum_{k=4}^\infty \frac{\partial^{2k-2}}{\partial(\beta h)^{2k-2}}
{\left. {\tilde{\chi}_0\left(\beta J_0^{(\Sigma)},\beta h\right)}
\right|_{_{\beta h =0} }
\frac{\left(\beta J^{(\Sigma)}\right)^{2k}}{(2k)!}}.
\end{eqnarray}

We recall that the
$k-2$-th derivative of $\tilde{\chi}_0\left(\beta J_0^{(\Sigma)},\beta h\right)$
with respect to the second argument, 
calculated at $h=0$, gives the total sum of all the $k$-th
cumulants normalized to $N$:
$\partial_{\beta h}^{k-2}\tilde{\chi}_0
\left(\beta J_0^{(\Sigma)},\beta h\right)|_{h=0}=\sum_{i_1,\ldots,i_k}
\media{\sigma_{i_1}\cdots\sigma_{i_k}}_0^{(c)}/N$, 
where $\media{\sigma_{i_1}\cdots\sigma_{i_k}}_0^{(c)}$ stands for the
cumulant, or connected correlation function, of order $k$ of the unperturbed
model, $\media{\sigma_{i_1}\sigma_{i_2}}_0^{(c)}=\media{\sigma_{i_1}\sigma_{i_2}}_0
-\media{\sigma_{i_1}}_0\media{\sigma_{i_2}}_0$, etc..
Note that, apart from the sign, 
these terms are proportional to the Binder cumulants \cite{Binder}
(which are all zero above $T_{c0}$ for $k>2$) only for $N$ finite. In the thermodynamic
limit the terms $b^{(\Sigma)}$, $c^{(\Sigma)}$, $\ldots$, 
in general are non zero 
and take into account the large deviations of the block-spin
distribution functions from the gaussian distribution.

Let $T_c^{(\Sigma)}=1/\beta_c^{(\Sigma)}$ be
the critical temperatures, if any, of the random model and 
let $t^{(\Sigma)}$ be the corresponding reduced temperatures:
\begin{eqnarray}
\label{THEOtt}
t^{(\Sigma)}\equiv \frac{T-T_c^{(\Sigma)}}{T_c^{(\Sigma)}}=
\frac{\beta_c^{(\Sigma)}-\beta}{\beta_c^{(\Sigma)}}+\mathop{O}(t^{(\Sigma)})^2.
\end{eqnarray} 
Here, the term ``critical temperature'', stands for any
temperature where some singularity shows up. However, if we limit ourself 
to consider only the critical temperatures crossing which the system passes from a P
region to a non P region, from Eq. (\ref{THEOO}) it is easy to see that,
independently on the sign of $J_0$ and on the nature of the phase transition,
we have the important inequalities
\begin{eqnarray}
\label{THEOf1}
\beta_c^{(\Sigma)}<\beta_{c0}^{(\Sigma)},
\end{eqnarray} 
where we have introduced $\beta_{c0}^{(\Sigma)}$, 
the inverse critical temperature of the unperturbed model
with coupling $J_0^{(\Sigma)}$ and zero external field.
If more than one critical temperature is present in the unperturbed model,
$\beta_{c0}^{(\Sigma)}$ is the value corresponding to the smallest value
of these critical temperatures (highest in terms of $\beta$).
Formally we set $\beta_{c0}^{(\Sigma)}=\infty$  
if no phase transition is present in the unperturbed model.
A consequence of Eq. (\ref{THEOL5}) is that, in studying the critical behavior
of the system for $h=0$, we can put $\tilde{h}^{(\Sigma)}=0$.
Throughout this paper, we shall reserve the name
critical temperature of the unperturbed model as a P-F
critical temperature through which the magnetization 
$m_0\left(\beta J_0,0\right)$ passes from a zero to a non zero value,
continuously or not. This implies, in particular, that for
$J_0<0$ we have - formally - $\beta_{c0}=\infty$. 

In this paper we shall study only the  
order parameters $m^{(\mathrm{F})}$ and $m^{(\mathrm{SG})}$,
whereas we will give only few remarks on how to
generalize the method for possible antiferromagnetic order parameters.
We point out however that the existence of possible antiferromagnetic transitions
of the unperturbed model does not affect the results we present
in this paper.

It is convenient to distinguish the cases $J_0\geq 0$ and $J_0<0$,
since they give rise to two strictly different scenarios.

\subsubsection{The case $J_0\geq 0$}
In this case $\beta J^{(\Sigma)}\tilde{\chi}_0 
\left(\beta J_0,\beta J^{(\Sigma)}m^{(\Sigma)} + \beta h\right)$
is an increasing function of $\beta$ 
for $\beta<\beta_{c0}^{(\Sigma)}$ (and for $h\geq 0$). As a consequence, 
we have that for sufficiently low temperatures,
the solution $m^{(\Sigma)}=0$ of Eq. (\ref{THEOa}) becomes 
unstable and two - and only two - non zero solutions 
$\pm m^{(\Sigma)}$ are instead favored.
The inverse critical temperatures 
$\beta_c^{(\mathrm{F})}$ and $\beta_c^{(\mathrm{SG})}$ can be determined
by developing - for $h=0$ - Eqs. (\ref{THEOa}) for small 
$m^{(\mathrm{F})}$ and $m^{(\mathrm{SG})}$, respectively, which, in terms of
$\tilde{\chi}_0$ gives the following exact equation
\begin{eqnarray}
\label{THEOg}
{\tilde{\chi}_0\left(\beta_c^{(\Sigma)}J_0^{(\Sigma)},0\right)}
\beta_c^{(\Sigma)}J^{(\Sigma)}=1, \quad \beta_{c}^{(\Sigma)}<\beta_{c0}^{(\Sigma)},
\end{eqnarray} 
where the constrain $\beta_{c}^{(\Sigma)}<\beta_{c0}^{(\Sigma)}$ excludes other
possible spurious solutions that may appear when $d_0\geq 2$.

The critical behavior of the system can be derived by developing 
Eqs. (\ref{THEOa}) for small fields. Alternatively, one can study the
critical behavior by analyzing the Landau free energy density
$\psi^{(\Sigma)}(m^{(\Sigma)})$ given by Eq. (\ref{THEOL1}).

In the following we will suppose that for $J_0>0$, 
$b^{(\Sigma)}$ be positive (we have checked this hypothesis in all the models
we have until now considered and that will be reported in the forthcoming
part II of the work).
Furthermore, even if the sign of $c^{(\Sigma)}$   
cannot be in general \textit{a priori} established, for the convexity
of the function $f_0$ with respect to $\beta h$, the sum of the six-th term with
$\Delta \left(\beta f_0\right) \left(\beta J_0^{(\Sigma)},\beta
J^{(\Sigma)}m^{(\Sigma)}\right)$, in Eq. (\ref{THEOL1}) must go 
necessarily to $+\infty$ for $m^{(\Sigma)}\to\infty$. 
In conclusion, when $J_0\geq 0$, for the  critical
behavior of the system, the only relevant parameters of $\psi^{(\Sigma)}$ are 
$a^{(\Sigma)}$, $b^{(\Sigma)}$ and
$h^{(\Sigma)}=\tilde{\chi}(\beta J_0^{(\Sigma)},0)J^{(\Sigma)}h$, 
so that the critical behavior
can be immediately derived as in the Landau theory for the so called \textit{$m^4$ model}.
On noting that
\begin{eqnarray}
\label{THEOAa}
\left\{
\begin{array}{l}
a^{(\Sigma)}\geq 0, \qquad \mathrm{for} ~ t^{(\Sigma)}\geq 0,\\
a^{(\Sigma)}<0, \qquad \mathrm{for} ~ t^{(\Sigma)}<0,\\
\end{array}
\right.
\end{eqnarray}
it is convenient to define
\begin{eqnarray}
\label{THEOA}
A^{(\Sigma)}\equiv -\beta\frac{\partial}{\partial\beta}a^{(\Sigma)},
\end{eqnarray} 
so that we have
\begin{eqnarray}
\label{THEOA1}
a^{(\Sigma)}=A^{(\Sigma)}|_{_{\beta=\beta_c^{(\Sigma)}}}t^{(\Sigma)}
+\mathop{O}(t^{(\Sigma)})^2.
\end{eqnarray} 

Note that, due to the fact that $J_0\geq 0$, $A^{(\Sigma)}>0$,
and, as already mentioned, $b^{(\Sigma)}\geq 0$ as well. 
By using Eq. (\ref{THEOA1}) for $\beta<\beta_{c0}$ and near
$\beta_c^{(\Sigma)}$, we see that the minimum $\bar{m}^{(\Sigma)}$ 
of $\psi^{(\Sigma)}$, \textit{i.e.}, the solution
of Eq. (\ref{THEOa}) near the critical temperature, is given by 
\begin{eqnarray}
\label{THEOmag}
\bar{m}^{(\Sigma)}=\left\{
\begin{array}{l}
0,\quad \qquad \qquad \qquad \qquad \qquad \qquad t^{(\Sigma)}~\geq 0, \\
\sqrt{-\frac{A^{(\Sigma)}}
{b^{(\Sigma)}}
|_{_{\beta=\beta_c^{(\Sigma)}}}
t^{(\Sigma)}~}+\mathop{O}(t^{(\Sigma)}),
\quad t^{(\Sigma)}< 0.
\end{array}
\right.
\end{eqnarray} 

Similarly, we can write general formulas for the susceptibility and
the equation of state. We have
\begin{eqnarray}
\label{THEOchi}
\tilde{\chi}^{(\Sigma)}=\left\{
\begin{array}{l}
\frac{\beta J^{(\Sigma)}\tilde{\chi}_0\left( \beta J_0^{(\Sigma)},0\right)}
{A^{(\Sigma)}}|_{_{\beta=\beta_c^{(\Sigma)}}}
\frac{1}{t^{(\Sigma)}}+\mathop{O}(1),~ t^{(\Sigma)}\geq 0, \\
\frac{\beta J^{(\Sigma)}\tilde{\chi}_0\left( \beta J_0^{(\Sigma)},0\right)}
{-2A^{(\Sigma)}}|_{_{\beta=\beta_c^{(\Sigma)}}}
\frac{1}{t^{(\Sigma)}}+\mathop{O}(1),~ t^{(\Sigma)}<0,
\end{array}
\right.
\end{eqnarray} 
\begin{eqnarray}
\label{THEOstate}
\bar{m}^{(\Sigma)}(h)=\left[
\frac{\beta J^{(\Sigma)}\tilde{\chi}_0\left( \beta J_0^{(\Sigma)},0\right)}
{A^{(\Sigma)}}
\right]^{\frac{1}{3}}_{\beta=\beta_c^{(\Sigma)}} h^{\frac{1}{3}}
+\mathop{O}\left(h^{\frac{2}{3}}\right).
\end{eqnarray} 

Finally, on using Eqs. (\ref{THEOL1}) and (\ref{THEOmag}) we get
that the specific heat $\mathcal{C}^{(\Sigma)}$ 
has the following finite jump discontinuity at $\beta_c^{(\Sigma)}$
\begin{eqnarray}
\label{THEOheat}
\mathcal{C}^{(\Sigma)}=\left\{
\begin{array}{l}
\mathcal{C}_c^{(\Sigma)}, \qquad \qquad \qquad t^{(\Sigma)}~\geq 0, \\
\mathcal{C}_c^{(\Sigma)}+ \frac{\left(A^{(\Sigma)}\right)^2}{2b^{(\Sigma)}}
|_{_{\beta=\beta_c^{(\Sigma)}}},
~ t^{(\Sigma)}< 0,
\end{array}
\right.
\end{eqnarray} 
where $\mathcal{C}_c^{(\Sigma)}$ is the continuous part of the specific heat corresponding
to the part of the free energy density without $\psi^{(\Sigma)}$. 

Hence, as a very general result, 
independently of the underlying dimension $d_0$ and the 
added random connectivity $c$, provided positive,
we recover that 
the random model has always a mean-field critical behavior 
with a second order phase transition with the classical exponents
$\beta=1/2$, $\gamma=\gamma'=1$, $\delta=3$ and $\alpha=\alpha'=0$, and 
certain constant coefficients depending on the susceptibility $\tilde{\chi}_0$ 
and its derivatives calculated at $\beta=\beta_c^{(\Sigma)}$ and external field $h=0$. 
Note however, that the correlation length of the system 
calculated along the distance of $\mathcal{L}_0$, $||\cdot||_0$, remains finite also 
at $\beta_c^{(\Sigma)}$. In fact, from Eq. (\ref{THEOh}), for the two point correlation
function at distance $r\equiv ||i-j||_{_{_0}}$ in $\mathcal{L}_0$ we have
\begin{eqnarray}
\label{THEOCORR}
{{C}}^{(\Sigma)}(r)=
{{C}}_0(\beta J_0^{(\Sigma)},\beta J^{(\Sigma)}m^{(\Sigma)}+\beta h ;r).
\end{eqnarray} 
If we now assume for ${{C}}_0(\beta J_0,0;r)$ 
the following general Ornstein-Zernike form 
\begin{eqnarray}
\label{THEOCORR1}
{{C}}_0(\beta J_0,0;r)=\frac{e^{-r/\xi_0}}{f_0(r)},
\end{eqnarray} 
$f_0(r)=f_0(\beta J_0;r)$ being
a smooth function of $r$ (which has not to be confused with the free energy density), 
and $\xi_0=\xi_0(\beta J_0)$ the correlation length,
which is supposed to diverge only at $\beta_{c0}$ (if any),
on comparing Eqs. (\ref{THEOCORR}) and (\ref{THEOCORR1}) 
for $\beta\geq \beta_c^{(\Sigma)}$ we have
\begin{eqnarray}
\label{THEOCORR2}
{{C}}^{(\Sigma)}(r)=\frac{e^{-r/\xi^{(\Sigma)}}}{f^{(\Sigma)}(r)},
\end{eqnarray} 
where 
\begin{eqnarray}
\label{THEOCORR3}
f^{(\Sigma)}(r)=f_0(\beta J_0^{(\Sigma)};r),
\end{eqnarray} 
and 
\begin{eqnarray}
\label{THEOCORR4}
\xi^{(\Sigma)}=\xi_0(\beta J_0^{(\Sigma)}).
\end{eqnarray} 
Therefore, due to the inequalities (\ref{THEOf1}), we see that
\begin{eqnarray}
\label{THEOCORR5}
\xi^{(\Sigma)}|_{\beta=\beta_c^{(\Sigma)}}=
\xi_0(\beta_c^{(\Sigma)} J_0^{(\Sigma)})<\infty.
\end{eqnarray} 
The knowledge of ${{C}}_0(\beta J_0,\beta h;r)$ also for $h\neq 0$ would allows
us to find the general expression for ${{C}}^{(\Sigma)}(r)$ through 
Eq. (\ref{THEOCORR}) also for 
$\beta> \beta_c^{(\Sigma)}$. However, since 
${{C}}_0(\beta J_0,\beta h;r)$ has no critical behavior for $h\neq 0$, 
it follows that ${{C}}^{(\Sigma)}(r)$ cannot have a critical behavior 
for $\beta> \beta_c^{(\Sigma)}$ either 
(and then also for $\beta\to \beta_c^{(\Sigma)}$ from the right).
This result is consistent with \cite{Barrat}.

\subsubsection{The case $J_0<0$}
In this case $J_0^{\mathrm{(F)}}<0$, so that - in general - $\beta J^{\mathrm{(F)}}\tilde{\chi}_0 
\left(\beta J_0,\beta J^{\mathrm{(F)}}m^{\mathrm{(F)}}\right)$
is no longer a monotonic function of $\beta$. However, it is easy to see that
that for $\beta=0$ and $\beta\to\infty$, this function goes to 0.
Therefore, for a sufficiently large connectivity $c$, 
from Eq. (\ref{THEOO}) we see that it may there appear
at least two regions where the paramagnetic solution $m^{\mathrm{(F)}}=0$ is stable,
separated by a third region in which a non zero solution is instead stable.
However the situation is even more complicated since, unlike
the case $J_0\geq 0$, the non monotonicity of $\beta J^{\mathrm{(F)}}\tilde{\chi}_0 
\left(\beta J_0,\beta J^\mathrm{(F)}m^\mathrm{(F)}\right)$ reflects also
in the fact that the self-consistent Eq. (\ref{THEOa}) for $\Sigma=$F may have
more solutions of the kind $\pm m^\mathrm{(F)},\pm m'^\mathrm{(F)},\ldots$
which are still stable with respect to the stability condition (\ref{THEOO}),
for $h=0$. We face in fact here the problem to compare more stable solutions.
According to Eq. (\ref{THEOlead}), 
in the thermodynamic limit, among all the possible stable solutions,
only $\bar{m}^\mathrm{(F)}$, the solution that minimizes 
$L^\mathrm{(F)}$, survives, 
whereas the not leading ones play the role of metastable states. 
This kind of scenario, which includes also finite jump discontinuities,
has been besides observed in the context of 
small-world neural networks in \cite{Skantos} where
we even observe some analogy in the used formalism, at least 
for the simplest case of one binary pattern.

From Eqs. (\ref{THEOL3}) and (\ref{THEOL4}) we see that the signs of
the Landau coefficients $a^{(\Sigma)}$, $b^{(\Sigma)}$, 
$c^{(\Sigma)}$, $\ldots$, are functions of $\beta$ and $J_0$ only.
Given $J_0<0$, the most important quantity that features the non monotonicity of 
$\beta J^\mathrm{(F)}\tilde{\chi}_0 
\left(\beta J_0,\beta J^\mathrm{(F)}m^\mathrm{(F)}\right)$ 
is the minimum value of $\beta$ over which $b^\mathrm{(F)}$ becomes
negative:
\begin{eqnarray}
\label{THEOneg1}
b^\mathrm{(F)}\leq 0, \qquad \beta\geq\beta_*^\mathrm{(F)},
\end{eqnarray} 
or, in terms of temperatures
\begin{eqnarray}
\label{THEOneg1b}
b^\mathrm{(F)}\leq 0, \qquad T\leq T_*^\mathrm{(F)}.
\end{eqnarray} 

The equation for $\beta_*^\mathrm{(F)}$, as a function of $J_0$,
defines a point  where $b^\mathrm{(F)}=0$. 
If $J_0<0$, the most general equation for a generic
critical temperature is no longer given by Eq. (\ref{THEOg}).
In fact, in general, a critical temperature now is any temperature where
the stable and leading solution $\bar{m}^\mathrm{(F)}$ 
may have a singular behavior, also with finite jumps between two non zero values. 

There are some simplification when for the Landau coefficient $c^\mathrm{(F)}$,
we have $c^\mathrm{(F)}>0$,
or at least $c^\mathrm{(F)}>0$ out of the P region
(this is the case of a model that we will report  
in the part II of the work). In this situation in
fact, from Eq. (\ref{THEOL1}) we see that $a^\mathrm{(F)}$,
$b^\mathrm{(F)}$ and $c^\mathrm{(F)}$ are the only relevant terms
for the critical behavior of the system and 
- for small values of $\bar{m}^\mathrm{(F)}$ -
we can again apply the Landau
theory, this time for the so called \textit{$m^6$ model}. In such a case, 
for the solution $\bar{m}^\mathrm{(F)}$ we have
\begin{eqnarray}
\label{THEOneg2}
\bar{m}^\mathrm{(F)}&=&
\sqrt{\frac{1}{2c^\mathrm{(F)}}\left(\sqrt{\left(b^\mathrm{(F)}\right)^2
-4a^\mathrm{(F)}c^\mathrm{(F)}}-b^\mathrm{(F)}\right)}, \quad \mathrm{if}, 
\nonumber \\
&& a^\mathrm{(F)}<0, \quad \mathrm{or} 
\nonumber \\
&& a^\mathrm{(F)}\geq 0 \quad \mathrm{and} \quad
b^\mathrm{(F)}\leq -4\sqrt{\frac{a^\mathrm{(F)}c^\mathrm{(F)}}{3}}, 
\end{eqnarray} 
whereas 
\begin{eqnarray}
\label{THEOneg3}
\bar{m}^\mathrm{(F)}&=& 0, \quad \mathrm{if},
\nonumber \\
&& a^\mathrm{(F)}\geq 0 \quad \mathrm{and} \quad 
b^\mathrm{(F)}> -4\sqrt{\frac{a^\mathrm{(F)}c^\mathrm{(F)}}{3}}.
\end{eqnarray} 
From Eqs. (\ref{THEOneg2}) and (\ref{THEOneg3}) 
we see that, if $b^\mathrm{(F)}>0$, 
we have a second order phase transition and 
Eqs. (\ref{THEOg})-(\ref{THEOCORR5}) are recovered with
Eq. (\ref{THEOneg2}) becoming the second of Eqs. (\ref{THEOmag})
for small and negative values of $a^\mathrm{(F)}$. 
However, from Eq. (\ref{THEOneg2}) we see that,
if $b^\mathrm{(F)}$ is sufficiently negative, 
we have a first order phase transition which, 
for small values of $a^\mathrm{(F)}$, gives
\begin{eqnarray}
\label{THEOneg4}
\bar{m}^\mathrm{(F)}&=& 
\sqrt{-\frac{b^\mathrm{(F)}}{c^\mathrm{(F)}}}
\left(1-\frac{a^\mathrm{(F)}c^\mathrm{(F)}}{2\left(b^\mathrm{(F)}\right)^2}\right),
\quad \mathrm{if},
\nonumber \\
&& a^\mathrm{(F)}<0 \quad \mathrm{and} \quad b^\mathrm{(F)}<0,\quad \mathrm{or} 
\nonumber \\
&& a^\mathrm{(F)}\geq 0 \quad \mathrm{and} \quad
b^\mathrm{(F)}\leq -4\sqrt{\frac{a^\mathrm{(F)}c^\mathrm{(F)}}{3}}.
\end{eqnarray} 
From Eq. (\ref{THEOneg2}) we see that the line $b^\mathrm{(F)}=-4\sqrt{a^\mathrm{(F)}c^\mathrm{(F)}/3}$
with $a^\mathrm{(F)}\geq 0$ establishes a line of first order transitions over which 
$\bar{m}^\mathrm{(F)}$ changes discontinuously from zero to 
\begin{eqnarray}
\label{THEOneg5}
\Delta\bar{m}^\mathrm{(F)}= 
\left(\frac{3a^\mathrm{(F)}}{c^\mathrm{(F)}}\right)^{\frac{1}{4}}.
\end{eqnarray} 
The point $a^\mathrm{(F)}=b^\mathrm{(F)}=0$ is a tricritical point where 
the second and first order transition lines meet. If we approach the
tricritical point along the line $b^\mathrm{(F)}=0$ we get 
the critical indices $\alpha=1/2$, $\alpha'=0$, $\beta=1/4$, $\gamma=\gamma'=1$ and
$\delta=5$. However, this critical behavior along the line
$b^\mathrm{(F)}=0$ has not a great practical interest since from
Eq. (\ref{THEOL3}) we see that it is not possible to keep $b^\mathrm{(F)}$
constant and zero as the temperature varies. 
Finally, we point out that, even if $c^\mathrm{(F)}>0$, 
when the transition is of the first order,
Eqs. (\ref{THEOneg2}) and (\ref{THEOneg4}) hold only for $b^\mathrm{(F)}$,
and then $a^\mathrm{(F)}$, sufficiently small, since
only in such a case the finite discontinuity of $\bar{m}^\mathrm{(F)}$
is small and then the truncation of the Landau free energy term $\psi^\mathrm{(F)}$
to a finite order meaningful. Note that this question implies also
that we cannot establish a simple and general rule to determine
the critical temperature of a first order phase transition (we will return
soon on this point).

When $c^\mathrm{(F)}<0$, the Landau theory of the $m^6$ model
cannot be of course applied.
However, as in the case $J_0>0$, even if the sign of $c^\mathrm{(F)}$   
cannot be \textit{a priori} established, for the convexity
of the function $f_0$ with respect to $\beta h$, the sum of the six-th term with
$\Delta \left(\beta f_0\right) \left(\beta J_0^\mathrm{(F)},\beta
J^\mathrm{(F)}m^\mathrm{(F)}\right)$, in Eq. (\ref{THEOL1}) must go 
necessarily to $+\infty$ for $m^\mathrm{(F)}\to\infty$ 
and a qualitative similar behavior of the $m^6$ model
is expected. 
In general, when $J_0<0$, the exact results are limited to the following ones. 

From now on, if not otherwise explicitly said, we shall reserve the name critical temperature, 
whose inverse value of $\beta$ we still indicate with $\beta_c^\mathrm{(F)}$, 
to any temperature on the boundary of a P region (through which $\bar{m}^\mathrm{(F)}$
passes from 0 to a non zero value, continuously or not).
For each critical temperature, 
depending on the value of $\beta_*^\mathrm{(F)}$, we have three possible
scenario of phase transitions:

\begin{eqnarray}
\label{THEOneg7}
\left\{
\begin{array}{l}
\beta_{c}^\mathrm{(F)}>\beta_*^\mathrm{(F)} \quad \Rightarrow \mathrm{first~order},\\
\beta_{c}^\mathrm{(F)}=\beta_*^\mathrm{(F)} \quad \Rightarrow \mathrm{tricritical~point},\\
\beta_{c}^\mathrm{(F)}<\beta_*^\mathrm{(F)} \quad \Rightarrow \mathrm{second~order},
\end{array}
\right.
\end{eqnarray} 
or, in terms of temperatures,
\begin{eqnarray}
\label{THEOneg7b}
\left\{
\begin{array}{l}
T_{c}^\mathrm{(F)}<T_*^\mathrm{(F)} \quad \Rightarrow \mathrm{first~order},\\
T_{c}^\mathrm{(F)}=T_*^\mathrm{(F)} \quad \Rightarrow \mathrm{tricritical~point},\\
T_{c}^\mathrm{(F)}>T_*^\mathrm{(F)} \quad \Rightarrow \mathrm{second~order}.
\end{array}
\right.
\end{eqnarray} 

Note that, according to our definition of critical temperature, 
the critical behavior described by Eqs. (\ref{THEOneg2}) - (\ref{THEOneg5})
represents a particular case of the general scenario expressed by
Eqs. (\ref{THEOneg7}). We see also that, in general, for the critical exponent 
$\beta$ we have $\beta\leq 1/4$.

In the case in which $\beta_{c}^\mathrm{(F)}$ corresponds to
a second order phase transition, or in the case in which
$a^\mathrm{(F)}<0$ out of the P region (at least immediately near 
the critical temperature), $\beta_{c}^\mathrm{(F)}$ can be exactly
calculated by Eq. (\ref{THEOg}). When we are not in such cases, the only
exact way to determine the critical temperature is to find 
the full solution for $\bar{m}^\mathrm{(F)}$ which consists in
looking numerically 
for all the possible solutions of Eq. (\ref{THEOa}) and - among those satisfying
the stability condition (\ref{THEOO}) - selecting the one that gives the
minimum value of $L^\mathrm{(F)}$.

\subsection{Level of accuracy of the method}
In the P region, Eqs. (\ref{THEOa}-\ref{THEOC2b}) are exact, 
whereas in the other regions
provide an effective approximation whose level of accuracy depends  
on the details of the model. In particular, in the absence of frustration
the method becomes exact at any temperature in two important limits:
in the limit $c\to 0^+$, in the case of second-order phase transitions, due
to a simple continuity argument;
and in the limit $c\to\infty$, due to the fact that in this
case the system becomes a suitable fully connected model exactly described
by the self-consistent equations (\ref{THEOa}) (of course, 
when $c\to\infty$, to have a finite critical temperature one has to renormalize 
the average of the coupling by $c$).  

However, for any $c>0$, off of the P region
and infinitely near the critical temperature, Eqs. (\ref{THEOa}-\ref{THEOh}) 
are able to give the exact critical 
behavior in the sense of the critical indices and, in the limit
of low temperatures, Eqs. (\ref{THEOa}-\ref{THEOe}) provide the exact
percolation threshold. 
In general, as for the SK model, which can be seen as 
a particular model with $J_0=0$,
the level of accuracy is better for the F phase rather than for
the SG one and this is particularly true 
for the free energy density $f^{(\Sigma)}$, Eq. (\ref{THEOl}).
In fact, though the derivatives of $f^{(\Sigma)}$ are expected
to give a good qualitative and partly also a quantitative
description of the system,
$f^\mathrm{(SG)}$ itself can give wrong results when the SG phase
at low temperatures is considered. We warn the reader that
in a model with $J_0=0$, and a symmetrical distribution $d\mu(J_{i,j})$ with variance
$\tilde{J}$, the method gives a ground state energy per site $u^{(\mathrm{SG})}$,
which grows with $c$ as $u^{(\mathrm{SG})}\sim -\tilde{J}c$, whereas the correct
result is expected to be $u^{(\mathrm{SG})}\sim -\tilde{J} \sqrt{c}$.
As a consequence, in the SK model, 
in the limit $\beta\to\infty$, the method gives a completely
wrong result with an infinite energy. 
We stress however that the order parameters $m^{(\mathrm{F})}$ and $m^{(\mathrm{SG})}$,
and then also the correlation functions, by construction, 
are exact in the zero temperature limit. 

\subsection{Phase diagram}
The inverse critical temperature $\beta_c$ of the random model is in general
a non single-value function of $\bm{X}$: $\beta_c=\beta_c(\bm{X})$, where
$\bm{X}$ represents symbolically the parameters of the
probability $d\mu$ for the couplings $J_{i,j}$, 
and the parameter $c$, the average connectivity (which is also a parameter of the 
probability distribution of the short-cut bonds). The  
parameters of $d\mu$ can be expressed through the moments of $d\mu$, and as they vary the
probability $d\mu$ changes. For example, if $d\mu$ is a Gaussian 
distribution, as in the SK model, there are only two parameters given by
the first and second moment.

In the thermodynamic limit, only one of the two solutions with label F or SG survives,
and it is the solution having minimum free energy.
In principle, were our method exact at all temperatures, we were able to derive
exactly all the phase diagram. 
However, in our method, the solution with label F or SG are exact only in their own P region, \textit{i.e.},
the region where $m^{(\mathrm{F})}=0$ or $m^{(\mathrm{SG})}=0$, respectively. 
Unfortunately, according to what we have seen in Sec. IIIC, 
whereas the solution with label F is still a good approximation also out of the P region,
in frustrated model (where the variance of $d\mu$ is large if compared to
the first moment) the free energy of the solution with label SG becomes 
completely wrong at low temperatures.
In conclusion, therefore, we are not able to give in general the exact boundary between the solution
with label F and the solution with label SG, and in particular we are not able 
to give the frontier F/SG. 
However, within some limitations which we now prescribe, 
we are able to give the exact critical surface,
\textit{i.e.}, the boundary with the P phase,
establishing which one - in the thermodynamic limit - of the two
critical boundaries, P-F or P-SG, is stable (we will use here the more common expression ``stable''
instead of the expression ``leading''), and to localize
some regions of the phase diagram for which we can
say exactly whether the stable solution is P, F, or SG. 
We will prove the stability of these solutions in Sec. VI.
When for a region we are not able to discriminate between
the solution with label F and the solution with label SG
and they are both out of their own P region, we will indicate
such a region with the symbol ``SG and/or F'' (stressing in this way that in this region there may be also
mixed phases and re-entrance phenomena).

\begin{figure}
\epsfxsize=75mm \centerline{\epsffile{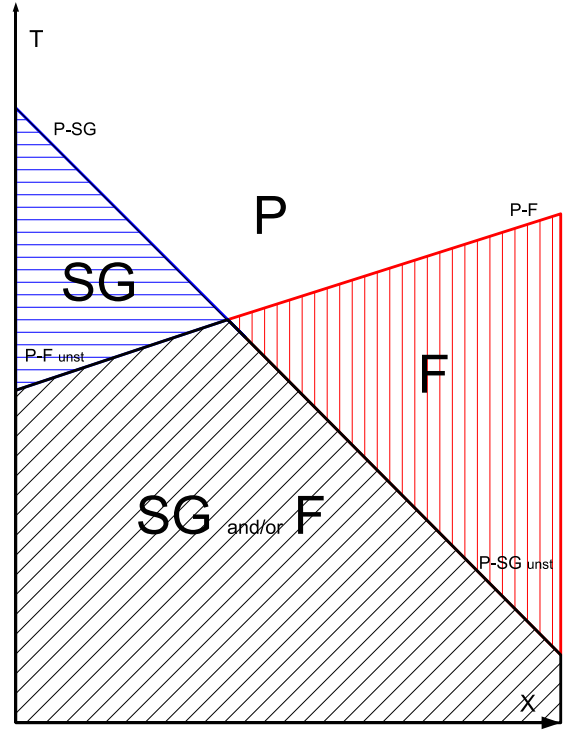}}
\caption{Phase diagram for the \textbf{case (1)}: $J_0\geq 0$ and $d_0<2$ or $d_0=\infty$ in a broad sense.} 
\label{phase_d1}
\end{figure}

\begin{figure}
\epsfxsize=75mm \centerline{\epsffile{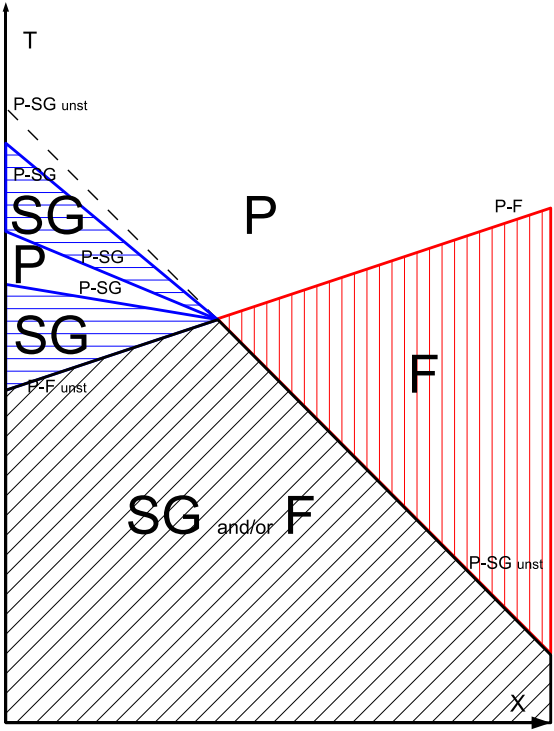}}
\caption{Phase diagram for the \textbf{case (2)}: $J_0\geq 0$ and $2\leq d_0<\infty$.} 
\label{phase_d2}
\end{figure}

\begin{figure}
\epsfxsize=75mm \centerline{\epsffile{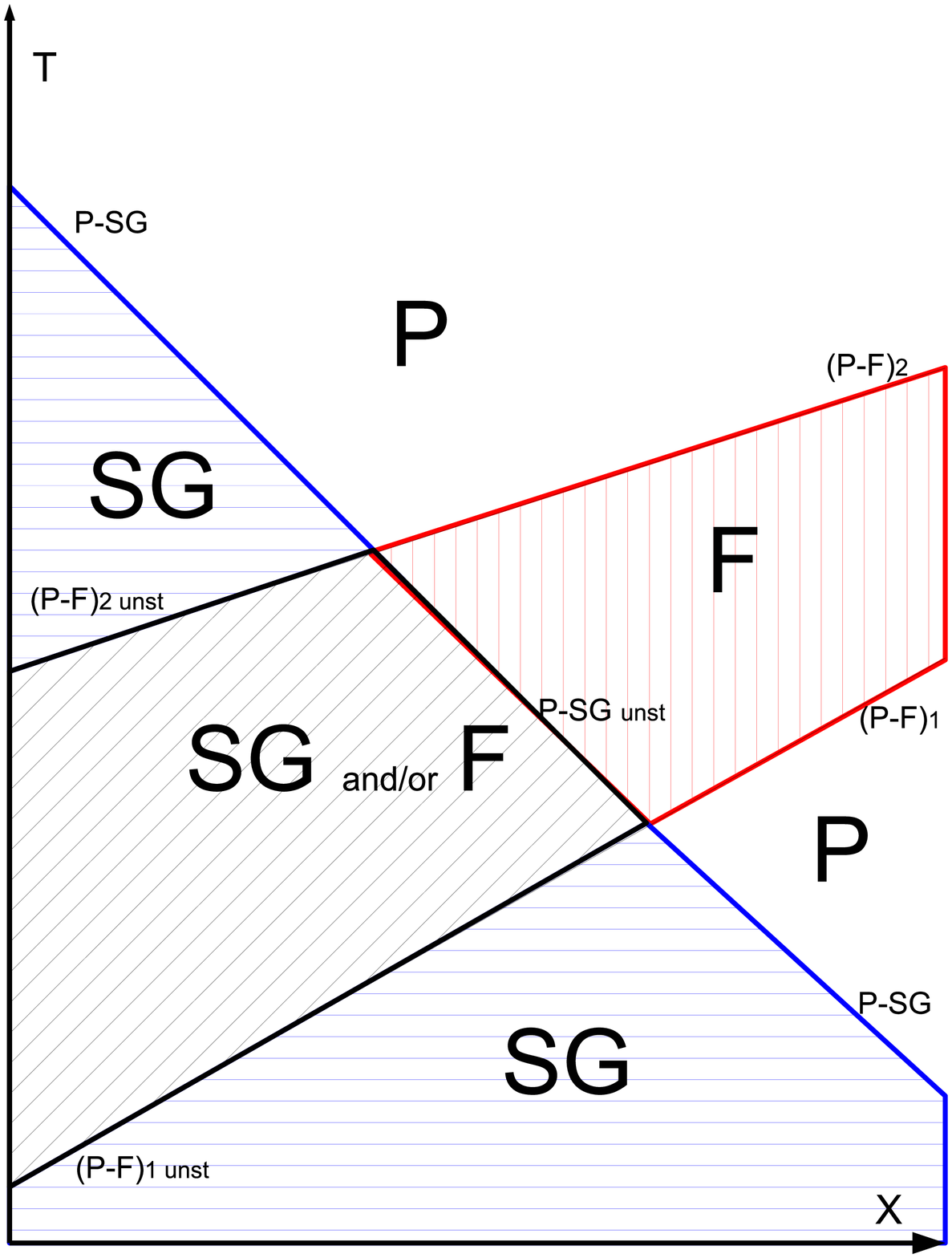}}
\caption{Phase diagram for the \textbf{case (3)}: $J_0< 0$ and $d_0<2$ or $d_0=\infty$ in a broad sense.} 
\label{phase_d3}
\end{figure}

\begin{figure}
\epsfxsize=75mm \centerline{\epsffile{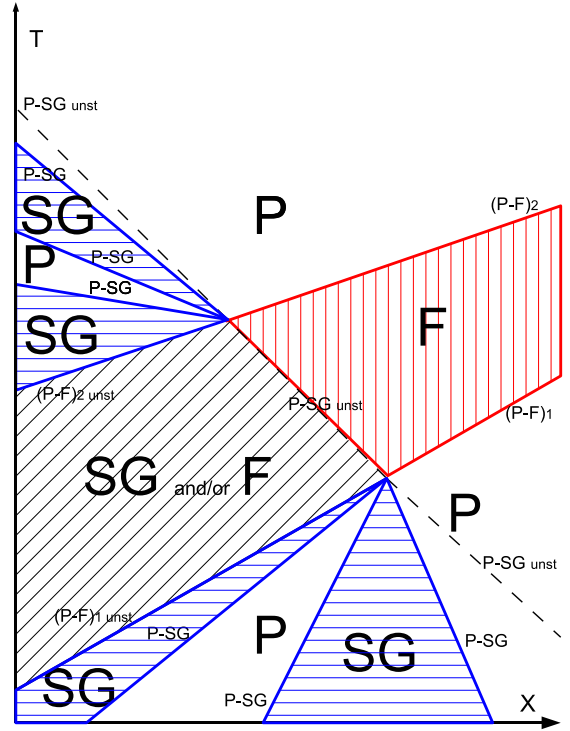}}
\caption{Phase diagram for the \textbf{case (4)}: $J_0< 0$ and $2\leq d_0<\infty$.} 
\label{phase_d4}
\end{figure}

In Sec. VI we prove that there are four possible kind of
phase diagrams that may occur according to the \textbf{cases}
\textbf{(1)} $(J_0\geq 0;~d_0<2,~\mathrm{or}~d_0=\infty)$, \textbf{(2)} $(J_0\geq 0;~2\leq d_0<\infty)$,
\textbf{(3)} $(J_0< 0;~d_0<2,~\mathrm{or}~d_0=\infty)$, and \textbf{(4)} $(J_0< 0;~2\leq d_0<\infty)$.  
The four kind of possible phase diagrams are schematically depicted 
in Figs. \ref{phase_d1}-\ref{phase_d4} in the plane $(T,~\bm{X})$.

\subsubsection{$J_0\geq 0$}
As we have seen in Sec. IIIB1, if $J_0\geq 0$, for both the solution with label F
and SG, we have one - and only one - critical temperature.

The stable inverse critical temperature $\beta_c$ satisfies 
the following rules.

\textbf{Case (1)}: If $d_0<2$ and $J_0$ is a finite range coupling, or else $d_0=\infty$ at least
in a broad sense (see \cite{MOIII}), $\beta_c(\bm{X})$ is a single-value function of $\bm{X}$, and we have
\begin{eqnarray}
\label{THEOrule}
\beta_c=\mathrm{min}\{\beta_c^{(\mathrm{F})},\beta_c^{(\mathrm{SG})}\},
\end{eqnarray} 
or, in terms of temperatures 
\begin{eqnarray}
\label{THEOrule1}
T_c=\mathrm{max}\{T_c^{(\mathrm{F})},T_c^{(\mathrm{SG})}\}.
\end{eqnarray} 
A schematic representation of this case is given in Fig.~1.

\textbf{Case (2)}: If instead $2\leq d_0<\infty$ we have 
\begin{eqnarray}
\label{THEOrule2}
\left\{
\begin{array}{l}
\beta_c=\beta_c^{(\mathrm{F})}, \quad \quad \quad \quad ~~ \mathrm{if} 
\quad \beta_c^{(\mathrm{SG})}\geq\beta_c^{(\mathrm{F})}, \\
\beta_c^{(\mathrm{F})}\geq \beta_c>\beta_c^{(\mathrm{SG})}, \quad \mathrm{if} 
\quad \beta_c^{(\mathrm{SG})}<\beta_c^{(\mathrm{F})},
\end{array}
\right.
\end{eqnarray} 
or, in terms of temperatures
\begin{eqnarray}
\label{THEOrule3}
\left\{
\begin{array}{l}
T_c=T_c^{(\mathrm{F})}, \quad \quad \quad \quad ~~ \mathrm{if} 
\quad T_c^{(\mathrm{SG})}\leq T_c^{(\mathrm{F})}, \\
T_c^{(\mathrm{F})}\leq T_c<T_c^{(\mathrm{SG})}, \quad \mathrm{if} 
\quad T_c^{(\mathrm{SG})}>T_c^{(\mathrm{F})}.
\end{array}
\right.
\end{eqnarray} 
Notice in particular that the second line of Eq. (\ref{THEOrule2}) (or Eq. (\ref{THEOrule3}))
does not exclude that $\beta_c(\bm{X})$ may be a non single-value function of $\bm{X}$. 
A schematic representation of this case is given in Fig. 2.

\subsubsection{$J_0< 0$}
As we have seen in Sec. IIIB2, if $J_0< 0$, for a sufficiently large connectivity $c$, 
the solution with label F has at least two separated P regions corresponding
to two critical temperatures. 
Here we assume that the underlying lattice $\mathcal{L}_0$
has only loops of even length so that, for example, triangular lattices
are here excluded. 
Let us suppose to have for the solution with label F only two critical temperatures 
(the minimum number, if $J_0<0$), and let be
\begin{eqnarray}
\label{THEOrule4}
\beta_{c1}^\mathrm{(F)}\geq \beta_{c2}^\mathrm{(F)},
\end{eqnarray} 
or, in terms of temperatures,
\begin{eqnarray}
\label{THEOrule5}
T_{c1}^\mathrm{(F)}\leq T_{c2}^\mathrm{(F)}.
\end{eqnarray} 

In general we have the following scenario.

\textbf{Case (3)}: If $d_0<2$ and $J_0$ is a finite range coupling, or $d_0=\infty$ in a broad
sense (see \cite{MOIII}), $\beta_{c2}(\bm{X})$ is a single-value function of $\bm{X}$
and satisfies Eq. (\ref{THEOrule}) 
(or, in terms of temperatures, Eq. (\ref{THEOrule1}) for $T_{c2}$).
The other critical inverse temperature $\beta_{c1}(\bm{X})$ is instead: 
either a two-value function of $\bm{X}$ and we have 
\begin{eqnarray}
\label{THEOrule6}
\beta_{c1}=\left(
\begin{array}{l}
\beta_{c1}^\mathrm{(F)} \\
\beta_{c}^\mathrm{(SG)}
\end{array}
\right), \quad \mathrm{if} \quad \beta_{c1}^\mathrm{(F)}\leq \beta_{c}^\mathrm{(SG)},
\end{eqnarray} 
or  
\begin{eqnarray}
\label{THEOrule7}
\nexists\quad \beta_{c1}, \quad \mathrm{if} \quad \beta_{c1}^\mathrm{(F)}> \beta_{c}^\mathrm{(SG)},
\end{eqnarray} 
where $\nexists$ in Eq. (\ref{THEOrule7}) means that if $\beta_{c1}^\mathrm{(F)}> \beta_{c}^\mathrm{(SG)}$
there is no stable boundary with the P region.     
A schematic representation of this case is given in Fig. 3.

\textbf{Case (4)}: If $2\leq d_0<\infty$, $\beta_{c2}$ satisfies Eq. (\ref{THEOrule2}) 
(or, in terms of temperature Eq. (\ref{THEOrule3}) for $T_{c2}$);
whereas for $\beta_{c1}$ we have either
\begin{eqnarray}
\label{THEOrule8}
\beta_{c1}=\left(
\begin{array}{l}
\beta_{c1}^\mathrm{(F)} \\
\beta_{c}^{\mathrm{(SG>)}}
\end{array}
\right), \quad \mathrm{if} \quad \beta_{c1}^\mathrm{(F)}\leq \beta_{c}^\mathrm{(SG)},
\end{eqnarray} 
or
\begin{eqnarray}
\label{THEOrule9}
\mathrm{if} \quad \exists \beta_{c1}\Rightarrow 
\beta_{c1}>\beta_{c}^{\mathrm{(SG)}}, \quad \mathrm{if} \quad \beta_{c1}^\mathrm{(F)}> \beta_{c}^\mathrm{(SG)},
\end{eqnarray} 
where in Eq. (\ref{THEOrule8}) we have introduced the symbol SG$>$ to indicate
that in general the stable P-SG surface is above (or below in terms of temperatures) the surface coming from the solution
with label SG:~$\beta_{c}^{\mathrm{(SG>)}}>\beta_{c}^\mathrm{(SG)}$.
Notice that, similarly to the case \textbf{(3)}, we cannot exclude that $\beta_{c1}$ in Eq. (\ref{THEOrule9})
be a non single-value function of $\bm{X}$, as well as $\beta_{c}^{\mathrm{(SG>)}}$ in Eq. (\ref{THEOrule8}).  
A schematic representation of this case is given in Fig.~4.

If more than two critical temperatures are present, the above scheme
generalizes straightforwardly.

Keeping our definition for the introduced symbol ``SG and/or F'',
we stress that: in all the fours cases the phases F and ``SG and/or F''
are exactly localized; in the cases \textbf{(1)} and \textbf{(3)} the phases P and SG are
exactly localized; in the cases \textbf{(2)} and \textbf{(4)} the SG phase is always 
limited below (in terms of temperatures) 
by the unstable P-SG surface coming from the solution with label SG 
(indicated as P-SG unst in Figs. 2 and 4). 
Finally, we stress that - under the hypothesis that $\mathcal{L}_0$ has only 
loops of even length - the stable P regions correspond always to the solution
with label F.

For $2\leq d_0<\infty$, from the second line of Eqs. (\ref{THEOrule2}) and (\ref{THEOrule8})
and from Eq. (\ref{THEOrule9}),
we see that the method is not able to give the complete information 
about the P-SG boundary since we have only inequalities,
not equalities. Furthermore, in these regions of the phase diagram
the critical temperature in general may be a non single-value function of $\bm{X}$.
On the other hand, we have the important information 
that in these equations the inequalities between $T_c$ and $T_c^{(\mathrm{SG})}$ 
(the solution with label SG) are always strict.
As a consequence, we see that, 
when $2\leq d_0<\infty$, in these regions
the SG ``magnetization'' $m^{(\mathrm{SG})}$
will always have a finite jump discontinuity in crossing the surface given by
$T_c$. In other words, along such a branch of the
critical surface corresponding to 
the second line of Eqs. (\ref{THEOrule3}) and (\ref{THEOrule8}) and Eq. (\ref{THEOrule9}),
we have a first order phase transition, independently of the fact that
the phase transition corresponding  to the $T_c^{(\mathrm{SG})}$ surface
is second-order, and independently on the 
sign of $J_0$. 


\subsection{Generalizations}
The generalization to the cases in which the unperturbed model
has an Hamiltonian $H_0$ involving couplings depending
on the bond $b\in\Gamma_0$ is straightforwardly. In this case 
we have just to substitute everywhere in the formulae 
(\ref{THEOa})-(\ref{THEOL6}), $J_0^{(\Sigma)}$ 
with the set $\{J_{0b}^{(\Sigma)}\}$. However, 
the critical behavior will be in general different and more complicated 
than that depicted in the Subsections IIIB1 and IIIB2.  
In particular, even in the case in which all the couplings
$J_{0b}^{(\Sigma)}$ are positive,
we cannot assume that the Landau coefficient $b^{(\Sigma)}$
be positive so that, even in such a case, first-order phase
transitions are in principle possible, as has been seen via MC simulations 
in undirected small-world models \cite{Sanchez}.

As anticipated, our method can be generalized also to study possible
antiferromagnetic phase transitions in the random model.
There can be two kind of sources of antiferromagnetism:
one due to a negative coupling $J_0$ in the unperturbed model,
the other due to random shortcuts $J_{i,j}$ having a measure $d\mu$ 
with a negative average. 

In the first case, if for example the sublattice $\mathcal{L}_0$ is bipartite 
into two sublattices $\mathcal{L}_0^{(a)}$ and $\mathcal{L}_0^{(b)}$,
the unperturbed model will have an antiferromagnetism described
by two fields $m_0^{(a)}$ and $m_0^{(b)}$. Correspondingly,
in the random model we will have to analyze two effective fields
$m^{(a)}$ and $m^{(b)}$ which will satisfy a set of two
coupled self-consistent equations similar to Eqs. (\ref{THEOa}) and
involving the knowledge of
$m_0^{(a)}$ and $m_0^{(b)}$. More in general, we can introduce the site-dependent 
solution $m_{0i}$ to find correspondingly in a set 
of coupled equations (at most $N$), the effective fields $m_i$
of the random model.

In the second case, following \cite{Almeida} we consider a lattice 
$\mathcal{L}_0$ which is composed of, say, $p$ sublattices
$\mathcal{L}_0^{(\nu)}$, $\nu=1,\ldots,p$. Then, we build up
the random model with the rule that any shortcut may connect
only sites belonging to two different sublattices. 
Hence, as already done in \cite{MOI} for the generalized SK model, we 
introduce $p$ effective fields $m^{(\nu)}$ which
satisfy a system of $p$ self-consistent equations involving
the $p$ fields 
$m_0^{(\nu)}$ and calculated in the $p$ external fields $J^{\mathrm{(F)}} m^{(\nu)}$
(note that here the symbol F stresses only the fact that the 
effective coupling must be calculated through Eq. (\ref{THEOb})).

\section{Mapping to non random models}
In Sec. V we will derive the main result presented in
Sec. III. To this aim in the next subsection we will recall the general mapping
between a random model, built up over a given graph, and a non random one
built up over the same graph, whereas in the following second subsection we will
generalize this mapping to random models built up over random graphs. 
We point out that the mapping does not consist in a sort of annealed approximation.

\subsection{Random Models defined on Quenched Graphs}
Let us consider the following random model.
Given a graph $\bm{g}$, which can be determined through the adjacency matrix
for shortness also indicate by $\bm{g}=\{g_b\}$, with $g_b=0,1$, $b$ being a bond,
let us indicate with $\Gamma_{\bm{g}}$ 
the set of the bonds $b$ of $\bm{g}$ and let us define over $\Gamma_{\bm{g}}$ 
the Hamiltonian 
\begin{eqnarray}
\label{HMD}
H\left(\{\sigma_i\};\{J_b\}\right)
\equiv -\sum_{b\in\Gamma_{\bm{g}}} J_b \sigma_{i_b}\sigma_{j_b} - \sum_i h_i \sigma_i
\end{eqnarray} 
where $J_b$ is the random coupling at the bond $b$, and 
$\sigma_{i_b},\sigma_{i_b}$ are the Ising variables at the end-points of $b$. 
The free energy $F$ and the physics are defined as in Sec. II by
Eqs. (\ref{logZ})-(\ref{OO}):
\begin{eqnarray}
\label{logZD}
-\beta F\equiv \int d\mathcal{P}\left(\{J_b\}\right)
\log\left(Z\left(\{J_b\}\right)\right),
\\
\label{logZD2}
\overline{\media{\mathcal{O}}^l}=\int d\mathcal{P}\left(\{J_b\}\right)
\media{\mathcal{O}}^l, \quad l=1,2
\end{eqnarray} 
%
where $d\mathcal{P}\left(\{J_b\}\right)$ 
is a product measure over all the possible bonds $b$ given 
in terms of normalized measures $d\mu_b\geq 0$ 
(we are considering a general measure $d\mu_b$ 
allowing also for a possible dependence on the bonds) 
\begin{eqnarray}
\label{dPM}
d\mathcal{P}\left(\{J_b\}\right)\equiv \prod_{b\in\Gamma_{\mathrm{full}}} 
d\mu_b\left( J_b \right),
\quad \int d\mu_b\left( J_b \right) =1,
\end{eqnarray}
where $\Gamma_{\mathrm{full}}$ stands for the set of bonds of the fully connected graph.
As in Sec. II, we will indicate a generic correlation function, connected or not, by
${{C}}$ with understood indices $i_1,\ldots,i_k$ all different, see
Eqs. (\ref{CF}) and (\ref{CG}).

In the following, given an arbitrary vertex $i$ of $\bm{g}$, 
we will consider as first neighbors $j$ of $i$ only those vertices 
for which $\int d\mu_{i,j}(J_{i,j})J_{i,j}$ or $\int d\mu_{i,j}(J_{i,j})J_{i,j}^2$ are at least $\mathop{O}(1/N)$.
Note that we can always neglect couplings having lower averages.  
We will indicate with $D(\Gamma_{\bm{g}})$ the average number of first
neighbors of the graph $\bm{g}$. For a $d$-dimensional lattice,
$D(\Gamma_{\bm{g}})=2d-1$, for a Bethe lattice of coordination number q, 
$D(\Gamma_{\bm{g}})=q-1$, and for long range models, $D(\Gamma_{\bm{g}})\propto N$.
We will exploit in particular the fact that 
$D(\Gamma_{\mathcal{L}_0}\cup\Gamma_{\mathrm{full}})=
D(\Gamma_{\mathrm{full}})\propto N$. 

Given a random model defined trough Eqs. (\ref{HMD}-\ref{dPM}), 
we define, on the same set of bonds $\Gamma_{\bm{g}}$, its \textit{related Ising model} 
trough the following Ising Hamiltonian
\begin{eqnarray}
\label{HI}
H_{I}\left(\{\sigma_i\};\{J_b^{(I)}\}\right)\equiv 
-\sum_{b\in\Gamma_{\bm{g}}} J_b^{(I)} \sigma_{i_b,j_b}
-\sum_i h_i \sigma_i,
\end{eqnarray} 
where the Ising couplings $J_b^{(I)}$ have 
non random values such that $~\forall ~b,b'\in \Gamma_{\bm{g}}$
\begin{eqnarray}
\label{JI}
J_{b'}^{(I)}&=&J_b^{(I)} \quad \mathrm{if} \quad 
d\mu_{b'}\equiv d\mu_{b}, \\
\label{JIb}
J_b^{(I)}&\neq & 0 
\quad \mathrm{if} \left\{
\begin{array}{l}
\quad \int d\mu_b(J_b)J_b=\mathop{O}\left(\frac{1}{N}\right), \quad \mathrm{or} \\
\quad \int d\mu_b(J_b)J_b^2=\mathop{O}\left(\frac{1}{N}\right). 
\end{array}
\right.
\end{eqnarray}
In the following a suffix $I$ over quantities such as $H_{I}$,
$F_{I}$, $f_{I}$, $g_I$, etc\ldots, or $J_b^{(I)}$, $\beta_c^{(I)}$, etc\ldots,
will be referred to the related Ising system with Hamiltonian (\ref{HI}).

We can always split the free energy of the random system with $N$ spins 
as follows
\begin{eqnarray}
\label{0logZ2}
-\beta F&=&\sum_{b} \int d\mu_{b} \log\left[\cosh(\beta J_b)\right]+ \nonumber \\
&& \sum_i \log\left[2\cosh(\beta h_i)\right] +\phi,
\end{eqnarray}
$\phi$ being the high temperature part of the free energy.
Let $\varphi$ be the density of $\phi$ in the thermodynamic limit
\begin{eqnarray}
\label{varphi}
\varphi \equiv \lim_{N\rightarrow \infty}\phi/N.
\end{eqnarray} 
Let us indicate with $\varphi_{I}$ the high temperature part
of the free energy density of the related Ising model defined through
Eqs. (\ref{HI}-\ref{JIb}). 
As is known, $\varphi_{I}$ can be expressed in terms of the
quantities $z_b=\tanh(\beta J_b^{(I)})$ and $z_i=\tanh(\beta h_i)$, \textit{i.e.}, 
the parameters of the high temperature expansion:
\begin{eqnarray}
\label{varphi1}
\varphi_{I} = \varphi_{I}\left(\{\tanh(\beta J_b^{(I)})\};\{\tanh(\beta h_i)\}\right).
\end{eqnarray} 

The related Ising model is defined by a set of,
typically few, independent couplings $\{J_b^{(I)}\}$, 
trough Eqs. (\ref{JI}-\ref{JIb})
and, for $h_i=0$, $i=1,\ldots,N$, its critical surface will be determined 
by the solutions of an equation, possibly vectorial, 
$G_I\left(\{\tanh(\beta J_b^{(I)})\}\right)=0$.

In \cite{MOI} we have proved the following mapping.

Let $\beta_c^{(\mathrm{SG})}$ and $\beta_c^{(\mathrm{F/AF})}$ be respectively 
solutions of the two equations 
\begin{eqnarray}
\label{mapp0g}
G_I\left(\{\int d\mu_b{\tanh^2(\beta_c^{(\mathrm{SG})} J_b)}\}\right)&=& 0, \\
\label{mapp01g}
G_I\left(\{\int d\mu_b{\tanh(\beta_c^{(\mathrm{F/AF})} J_b)}\}\right)&=& 0.
\end{eqnarray} 
Asymptotically, at sufficiently high dimensions $D(\Gamma_{\bm{g}})$,
the critical inverse temperature of the spin glass model $\beta_c$ is given by
\begin{eqnarray}
\label{mappg}
\beta_c=\mathrm{min}\{\beta_c^{(\mathrm{SG})},\beta_c^{(\mathrm{F/AF})}\};
\end{eqnarray} 
and in the paramagnetic phase for $D(\Gamma_{\bm{g}})>2$ the following mapping holds 
\begin{eqnarray}
\label{mapp1g}
\left|\frac{\varphi-\varphi_{eff}}{\varphi}\right|=
\left|\frac{{{C}}-{{C}}_{eff}}{{{C}}}\right|=
O\left(\frac{1}{D(\Gamma_{\bm{g}})}\right),
\end{eqnarray} 
\begin{eqnarray}
\label{mapp2g}
\varphi_{eff}=\frac{1}{l}\varphi^{(\Sigma)}\equiv\frac{1}{l}
\varphi_{I}\left(\{\int d\mu_b{\tanh^{l}(\beta J_b)}\}\right),
\end{eqnarray}
and
\begin{eqnarray}
\label{mapp2gc}
{{C}}_{eff}=\frac{1}{l}{{C}}^{(\Sigma)}\equiv\frac{1}{l}
{{C}}_{I}\left(\{\int d\mu_b{\tanh^{l}(\beta J_b)}\}\right),
\end{eqnarray}
where
\begin{eqnarray}
\label{mapp3g}
l=\left\{
\begin{array}{l}
2, \quad \mathrm{if}\quad
\varphi_{I}\left(\{\int d\mu_b{\tanh^{2}(\beta J_b)}\}\right)\geq \\
2|\varphi_{I}\left(\{\int d\mu_b{\tanh(\beta J_b)}\}\right)|, \\
1, \quad \mathrm{if}\quad
\varphi_{I}\left(\{\int d\mu_b{\tanh^{2}(\beta J_b)}\}\right)< \\
2|\varphi_{I}\left(\{\int d\mu_b{\tanh(\beta J_b)}\}\right)|,
\end{array}
\right.
\end{eqnarray} 
and $\Sigma$=F/AF or SG, for $l$=1 or 2, respectively.

In the limit $D(\Gamma_{\bm{g}})\rightarrow \infty$ and $h_i=0$, $i=1,\ldots,N$, 
Eqs. (\ref{mapp0g}-\ref{mapp3g}),
give the exact free energy and correlation functions 
in the paramagnetic phase (P); the
exact critical paramagnetic-spin glass (P-SG), $\beta_c^{(\mathrm{SG})}$, and
paramagnetic- F/AF (P-F/AF), 
$\beta_c^{(\mathrm{F/AF})}$, surfaces, whose reciprocal 
stability depends on which of the two ones has higher temperature. 
In the case of a measure $d\mu$ not depending on the bond $b$, 
the suffix F and AF stand
for ferromagnetic and antiferromagnetic, respectively.
In the general case, such a distinction is possible only
in the positive and negative sectors in the space of the 
parameters of the probability distribution, $\{\int d\mu_b J_b\geq 0\}$ and 
$\{\int d\mu_b J_b< 0\}$, respectively, whereas, for
the other sectors, we use the symbol
F/AF only to stress that the transition is not P-SG.

It is not difficult to see that, when the measure $d\mu$ 
does not depend on the specific bond $b$, \textit{i.e.}, if
$d\mu_b\equiv d\mu_{b'}~ \forall b,b'\in \Gamma_{\bm{g}}$, in the P region 
Eqs. (\ref{mapp0g}-\ref{mapp3g})
lead to the following exact limit for $\varphi$ and $C$ \cite{MOIII}
\begin{eqnarray}
\label{phi=0}
\lim_{D(\Gamma_{\bm{g}})\to \infty}\varphi=\lim_{D(\Gamma_{\bm{g}})\to \infty}{{C}}=0, 
\quad \mathrm{for}\quad \beta\leq\beta_c,
\end{eqnarray}
therefore, the basic role of Eqs. (\ref{mapp1g}-\ref{mapp3g}),
is to show how, in the limit $D(\Gamma_{\bm{g}})\to\infty$, $\varphi$ and ${{C}}$ 
approach zero and which are their singularities.
In particular this proves that for all the (random) infinite dimensional models
and any disorder non bond-dependent, the critical exponent $\alpha'$ for the specific
heat has the mean-field classical value, $\alpha'=0$, and that the correlation
functions (with different indices) above the critical temperature are exactly zero.
We point out however that, when the measure $d\mu_b$ depends explicitly on the
bond $b$, Eq. (\ref{phi=0}) in general does not hold~
\footnote{This was not strongly emphasized in \cite{MOIII}}.
In fact, when the measure $d\mu_b$ is bond-dependent, 
the symmetry expressed by Eq. (\ref{phi=0}) is broken since the bonds are no
longer equivalent. 
As we will see in the next section, in small-world models with an underlying 
lattice $\mathcal{L}_0$ having $d_0<2$, 
even if Eq. (\ref{phi=0}) may still holds for $\varphi$, the symmetry is
broken for ${{C}}$ since the direction(s) of the axis(es) of 
$\mathcal{L}_0$ is(are) now favored direction(s). Yet, 
if $2\leq d_0< \infty$, 
the symmetry (\ref{phi=0}) for $\varphi$ is broken as well.  

The analytic continuation of Eqs. (\ref{mapp1g}-\ref{mapp3g}) to
$\beta>\beta_c$ and/or for $h\neq 0$ provide 
certain estimations which are expected to 
be qualitatively good. 
In general such estimations are not exact,
and this is particularly evident for the free energy
density of the SG phase.
However, the analytic continuation for
the other quantities gives a good qualitative result and
provide the exact critical behavior (in the sense of the
critical indices) and the exact percolation threshold.

For models defined over graphs satisfying a weak definition
of infinite dimensionality, as happens on a Bethe lattice with coordination number
$q>2$, a more general mapping has been established \cite{MOIII}. In this case, all the
above equations - along the critical surface (at least) - 
still hold exactly in the thermodynamic limit, 
where we can set effectively $D(\Gamma_{\bm{g}})=\infty$.
However, for the aims of this paper we do not need here to consider
this generalization of the mapping.

We have yet to make an important comment about Eqs. (\ref{THEOa01}),
(\ref{THEOa02}) and (\ref{THEOa04}), concerning the evaluation
of a correlation function in the SG phase 
here for a random system with $J_0=0$ (for the moment being). 
In fact Eq. (\ref{mapp2gc}), for both a normal   
and a quadratic correlation function $C^{(1)}$ or $C^{(2)}$,
has a factor 1/2 not entering in the physical Eqs. (\ref{THEOa01}),
(\ref{THEOa02}) and (\ref{THEOa04}).
The difference is just due to an artefact of the mapping
that separates the Gibbs state into two pure states \cite{Parisi}
not only in the F case, but also in the SG case. 
In fact, let us consider the correlation functions
of order $k=1$, that is, $C^{(1)}=\overline{\media{\sigma_1}}$ and
$C^{(2)}=q_{EA}=\overline{\media{\sigma_1}^2}$. We see that, for $C^{(1)}$,
Eq. (\ref{mapp2gc}) in the SG phase gives $C^{(1)}=m^{\mathrm{(SG)}}/2$.
On the other hand, for any non zero solution $m^{\mathrm{(SG)}}$
of the self-consistent Eq. (\ref{THEOa}), there exists another solution
$-m^{\mathrm{(SG)}}$, and both the solutions have 1/2 probability
to be realized in the random model. Since the SG phase is expected
to be the phase characterized by having $q_{EA}\neq 0$ and
$\overline{\media{\sigma_1}}=0$, we see that if we introduce both the
solutions $m^{\mathrm{(SG)}}$ and $-m^{\mathrm{(SG)}}$, 
we get $\overline{\media{\sigma_1}}=0$ in the SG phase.
Similarly, for $C^{(2)}$,
Eq. (\ref{mapp2gc}) in the SG phase gives $C^{(2)}=(m^{\mathrm{(SG)}})^2/2$,
which at zero temperature gives 1/2, whereas a completely frozen 
state with $q_{EA}=1$ is expected. Again, we recover the expected physical $q_{EA}$
by using both the
solutions $m^{\mathrm{(SG)}}$ and $-m^{\mathrm{(SG)}}$.
Repeating a similar argument for any correlation function of order $k$,
and recalling that for $k$ even (odd) the correlation function is
an even (odd) function of the external magnetic field $h$, we arrive at 
Eqs. (\ref{THEOa01}), (\ref{THEOa02}) and (\ref{THEOa04}).

\subsection{Random Models defined on Unconstrained Random Graphs}
Let us consider now more general random models in which
the source of the randomness comes from both
the randomness of the couplings and the randomness of the graph.
Given an ensemble of graphs $\bm{g}\in\mathcal{G}$ distributed with
some distribution $P(\bm{g})$, let us define 
\begin{eqnarray}
\label{HM}
H_{\bm{g}}\left(\{\sigma_i\};\{J_b\}\right)
&\equiv& -\sum_{b\in\Gamma_{\bm{g}}} J_b \sigma_{i_b}\sigma_{j_b}
-h\sum_i\sigma_i\nonumber \\
&=& -\sum_{b\in\Gamma_{\mathrm{full}}} g_bJ_b \sigma_{i_b}\sigma_{j_b}
-h\sum_i\sigma_i.
\end{eqnarray} 
The free energy $F$ and the physics are now given by
\begin{eqnarray}
\label{logZDD}
-\beta F\equiv 
\sum_{{\bm{g}}\in\mathcal{G}}P({\bm{g}})\int d\mathcal{P}\left(\{J_b\}\right)
\log\left(Z_{\bm{g}}\left(\{J_b\}\right)\right),
\end{eqnarray} 
and similarly for $\overline{\media{O}}$.
Here $Z_g\left(\{J_b\}\right)$ 
is the partition function of the quenched system onto the graph realization 
$\bm{g}$ with bonds in $\Gamma_{\bm{g}}$ 
\begin{eqnarray}
\label{Z}
Z_{\bm{g}}\left(\{J_b\}\right)= \sum_{\{\sigma_i\}}e^{-\beta 
H_{\bm{g}}\left(\{\sigma_i\};\{J_b\}\right)}, 
\end{eqnarray} 
and $d\mathcal{P}\left(\{J_b\}\right)$ 
is again a product measure over all the possible bonds $b$ given 
as defined in Eq. (\ref{dPM}). Note that the bond-variables $\{g_b\}$ 
are independent from the coupling-variables $\{J_b\}$. 

For unconstrained random graphs, or for random graphs having a
number of constrains that grows sufficiently slowly with $N$, 
the probability $P(\bm{g})$, for large $N$, factorizes as
\begin{eqnarray}
P(\bm{g})=\prod_{b\in\Gamma_{\mathrm{full}}} p_b(g_b).
\end{eqnarray} 
In such a case we can exploit the mapping we have previously seen 
for models over quenched graphs
as follows.
Let us define the effective coupling $\tilde{J}_b$:
\begin{eqnarray}
\label{sep}
\tilde{J}_b\equiv J_b \cdot g_b, 
\end{eqnarray}
correspondingly, since the random variables $J_b$ and $g_b$
are independent we have
\begin{eqnarray}
\label{dmutilde}
d\tilde{\mu}_b(\tilde{J}_b)=d\mu_b(J_b) \cdot p_b(g_b), 
\end{eqnarray}
with the sum rule
\begin{eqnarray}
\int d\tilde{\mu}_b(\tilde{J}_b)f(J_b;g_b)=\sum_{g_b=0,1}p_b(g_b)\int d\mu_b(J_b)f(J_b;g_b).
\end{eqnarray}
As a consequence, if we define the following global measure
\begin{eqnarray}
d\tilde{\mathcal{P}}\left(\{\tilde{J}_b\}\right)=
P(\bm{g})\cdot d\mathcal{P}\left(\{J_b\}\right)=
\prod_{b\in\Gamma_{\mathrm{full}}}d\tilde{\mu}_b(\tilde{J}_b),
\end{eqnarray}
we see that
the mapping of the previous section can be applied as we had 
a single effective graph $\Gamma_p$ given by
\begin{eqnarray}
\label{GammaP}
\Gamma_p\equiv \{b\in\Gamma_{\mathrm{full}}:~p_b(g_b=1)\neq 0\},
\end{eqnarray} 
in fact we have
\begin{eqnarray}
\label{logZDD2}
-\beta F\equiv 
\int d\tilde{\mathcal{P}}\left(\{\tilde{J}_b\}\right)
\log\left(Z_p\left(\{\tilde{J}_b\}\right)\right),
\end{eqnarray}
where $Z_p$ is the partition function of the model with Hamiltonian $H_p$
given by
\begin{eqnarray}
\label{HM2}
H_p \left(\{\sigma_i\};\{\tilde{J}_b\}\right)
\equiv -\sum_{b\in\Gamma_p} \tilde{J}_b \sigma_{i_b}\sigma_{j_b}
-h\sum_i\sigma_i.
\end{eqnarray} 
%

\section{Derivation of the self-consistent equations}
By using the above results, we are now able to derive easily
Eqs. (\ref{THEOa}-\ref{THEOlead}). Sometimes to indicate
a bond $b$ we will use the symbol $(i,j)$, or more shortly $ij$.

It is convenient to look formally at the
coupling $J_0$ also as a random coupling with distribution
\begin{eqnarray} 
\label{dmu0}
d\mu_0(J_0')/dJ_0'=\delta (J_0'-J_0).
\end{eqnarray}

Let us rewrite explicitly the Hamiltonian (\ref{H}) as follows
\begin{eqnarray}
\label{H1}
H_{\bm{c}}&=&-\sum_{(i,j)\in\Gamma_0}\left(J_0+c_{ij}{J}_{ij}\right)\sigma_{i}\sigma_{j}
\nonumber \\
&& -\sum_{i<j,~(i,j)\notin \Gamma_0}c_{ij}{J}_{ij}\sigma_{i}\sigma_{j}
-h\sum_i\sigma_i,
\end{eqnarray}
and let us introduce the 
random variables $J_b'$, $g_b'$ and $\tilde{J}_b'$, where 
\begin{eqnarray}
J_b'\equiv  
\left\{
\begin{array}{l}
\label{sep2}
J_0+c_bJ_b, \quad b\in\Gamma_0,\\
\label{sep3}
J_b, \quad ~ b\notin\Gamma_0,
\end{array}
\right.
\end{eqnarray}
\begin{eqnarray}
g_b'\equiv  
\left\{
\begin{array}{l}
\label{sep2}
1, \quad b\in\Gamma_0,\\
\label{sep3}
c_b, \quad ~ b\notin\Gamma_0,
\end{array}
\right.
\end{eqnarray}
and 
\begin{eqnarray}
\label{sep1}
\tilde{J}_b'\equiv J_b' \cdot g_b'. 
\end{eqnarray}
Taking into account that 
the random variable $J_0+c_{ij}J_{ij}$, up to terms $\mathop{O}(1/N)$, is
distributed according to $d\mu_0(J_0)$,
the independent random variables $J_b'$ and $g_b'$ 
have distributions $d\mu_b'$ and $p_b'$ respectively given by
\begin{eqnarray}
d\mu_b'(J_b') = 
\left\{
\begin{array}{l}
\label{sep2}
d\mu_0(J_b'), \quad b\in\Gamma_0,\\
\label{sep3}
d\mu(J_b'), \quad ~ b\notin\Gamma_0,
\end{array}
\right.
\end{eqnarray}
and
\begin{eqnarray}
p_b'(g_b') = 
\left\{
\begin{array}{l}
\label{sep4}
\delta_{g_b',1}, \quad ~ b\in\Gamma_0 ,\\
\label{sep5}
p(g_b'), \quad b\notin\Gamma_0,
\end{array}
\right.
\end{eqnarray}
where the measures $d\mu$ and $p$ are those of the model introduced in Sec. II. 
As a consequence, Eq. (\ref{H1}) can be cast 
in the form of Eq. (\ref{HM2}) with the measure
\begin{eqnarray}
d\tilde{\mu}_b'(\tilde{J}_b') = 
\left\{
\begin{array}{l}
\label{sep2}
d\mu_0(J_b')\delta_{g_b',1}, \quad b\in\Gamma_0,\\
\label{sep3}
d\mu(J_b')p(g_b'), \quad ~ b\notin\Gamma_0.
\end{array}
\right.
\end{eqnarray}
Finally, since $p_b(g_b)\neq 0$ for any
$b\in\Gamma_{\mathrm{full}}$, we have also
\begin{eqnarray}
\Gamma_p=\Gamma_{\mathrm{full}},
\end{eqnarray} 
and due to the fact that $D(\Gamma_{\mathrm{full}})\propto N$, in the thermodynamic limit 
the mapping becomes exact. 

According to Eqs. (\ref{HI}-\ref{JIb}), the related Ising model
of our small-world model has the following Hamiltonian
with two free couplings: $J_0^{(I)}$, for $\Gamma_0$, and $J^{(I)}$, 
for $\Gamma_{\mathrm{full}}$
\begin{eqnarray}
\label{HR}
H_I&=&-J_0^{(I)}\sum_{(i,j)\in\Gamma_0}\sigma_{i}\sigma_{j}
-J^{(I)}\sum_{i<j,~(i,j)\notin \Gamma_0}\sigma_{i}\sigma_{j}
\nonumber \\
&&-h\sum_i\sigma_i.
\end{eqnarray}
After solving this Ising ($I$) model the mapping allows us to come back
to the random model by performing simultaneously for any $b\in \Gamma_{\mathrm{full}}$ 
the reverse substitutions
\begin{eqnarray}
\label{MT0}
\tanh\left(\beta J^{(I)}_b\right)\to \int
d\tilde{\mu}_b'(\tilde{J}_b')\left(\tilde{J}_b'\right)\tanh^l
\left(\beta \tilde{J}_b'\right), 
\end{eqnarray} 
where $l=1,2$ for $\Sigma=$ F or SG solution, respectively.
Since the couplings $J_0^{(I)}$ and $J^{(I)}$ are arbitrary, we find it convenient
to renormalize $J^{(I)}$ as $J^{(I)}/N$ and at the end of the 
calculation to put again $J^{(I)}$ instead of $J^{(I)}/N$.
Note that for the mapping 
nothing changes if we do not make this substitution;
the choice to use $J^{(I)}/N$ instead of $J^{(I)}$ is merely due
to a formal convenience, since in this way the calculations 
are presented in a more standard and physically understandable form. 
In fact, according to Eqs. (\ref{sep2}) and (\ref{MT0}) 
what matters after solving the related Ising model with 
$J^{(I)}/N$ instead of $J^{(I)}$
is that, once for $\Sigma$=F and once for $\Sigma$=SG,
we perform - simultaneously in the two couplings - 
the following reverse mapping transformations
($l=1,2$ for $\Sigma=$ F or SG, respectively):
\begin{eqnarray}
\label{MT}
&& \tanh\left(\beta J^{(I)}/N\right)\to \int
d\tilde{\mu}\left(\tilde{J}_{ij}\right)\tanh^l
\left(\beta \tilde{J}_{ij}\right),
\end{eqnarray}
for $(i,j)\notin \Gamma_0$, and
\begin{eqnarray}
\label{MT2}
&& \tanh\left(\beta J_0^{(I)}\right)\to \int
d\tilde{\mu}\left(\tilde{J}_{ij}\right)\tanh^l
\left(\beta \tilde{J}_{ij}\right),
\end{eqnarray}
for $(i,j)\in \Gamma_0$.
 
Explicitly, by applying Eqs. (\ref{sep2}) and Eq. (\ref{PP}) the transformations 
(\ref{MT}) and (\ref{MT2}) become, respectively
\begin{eqnarray}
\label{MT3}
\beta J^{(I)} \to \beta J^{(\Sigma)}
\end{eqnarray}
and
\begin{eqnarray}
\label{MT4}
\beta J_0^{(I)} \to \beta J_0^{(\Sigma)},
\end{eqnarray}
where we have made use of the definitions (\ref{THEOb})-(\ref{THEOe})
introduced in Sec. III.

Let us now solve the related Ising model.
We have to evaluate the following partition function
\begin{eqnarray}
\label{ZI}
Z_I=\sum_{\{\sigma_i\}}e^{\beta J_0^{(I)}\sum_{(i,j)\in\Gamma_0}\sigma_{i}\sigma_{j}+
\beta \frac{J^{(I)}}{2N}\sum_{i\neq j}\sigma_{i}\sigma_{j}
+\beta h\sum_i\sigma_i}.
\end{eqnarray}
In the following we will suppose that $J^{(I)}$ (and then $J^{(\Sigma)}$) is positive.
The derivation for $J^{(I)}$ (and then $J^{\mathrm{(F)}}$) negative differs from the other
derivation just for a rotation of $\pi/2$ in the complex $m$-plane, and leads to the
same result one can obtain by analytically continue the equations derived 
for $J^{(I)}>0$ to the region $J^{(I)}<0$. 

By using the Gaussian transformation we can rewrite $Z_I$ as
\begin{eqnarray}
\label{ZI1}
Z_I&=& c_N 
\sum_{\{\sigma_i\}}e^{\beta J_0^{(I)}\sum_{(i,j)\in\Gamma_0}\sigma_{i}\sigma_{j}}
\nonumber \\ 
&& \times \int_{-\infty}^{\infty} d{{m}} ~ e^{-\frac{\beta}{2} J^{(I)}{{m}}^2N +
\beta\left(J^{(I)}{{m}} + h\right)\sum_i\sigma_i},
\end{eqnarray}
where $c_N$ is a normalization constant
\begin{eqnarray}
c_N = \sqrt{\frac{\beta J^{(I)}N}{2\pi}},
\end{eqnarray}
and, in the exponent of Eq. (\ref{ZI1}), 
we have again neglected terms of order $\mathop{O}(1)$.
For finite $N$ we can exchange the integral and the sum over the $\sigma$'s.
By using the definition of the unperturbed model with Hamiltonian $H_0$,
Eq. (\ref{H0}), whose free energy density, for given $\beta J_0$ and $\beta h$,
is indicated with $f_0(\beta J_0,\beta h)$, we arrive at
\begin{eqnarray}
\label{ZI2a}
Z_I&=& c_N \int_{-\infty}^{\infty} d{{m}} ~ e^{-N L({{m}})},
\end{eqnarray}
where we have introduced the function 
\begin{eqnarray}
\label{ZI2}
L({{m}})= \frac{\beta}{2} J^{(I)}{{m}}^2 
+\beta f_0 \left(\beta J_0^{(I)},\beta J^{(I)}{{m}} + \beta h\right).
\end{eqnarray}
By using 
$\partial_{\beta h}~\beta f_0(\beta J_0,\beta h)=-m_0(\beta J_0,\beta h)$, and
$\partial_{\beta h}~m_0(\beta J_0,\beta h)=\tilde{\chi}_0(\beta J_0,\beta h)$
we get  
\begin{eqnarray}
\label{ZI3}
L'({{m}})= \beta J^{(I)}
\left[{{m}}-m_0 \left(\beta J_0^{(I)},\beta J^{(I)}{{m}} + \beta h\right)\right], 
\end{eqnarray}
\begin{eqnarray}
\label{ZI4}
L''({{m}})= \beta J^{(I)}\left[1- \beta J^{(I)} 
\tilde{\chi}_0 \left(\beta J_0^{(I)},\beta J^{(I)}{{m}} + \beta h\right)\right]. 
\end{eqnarray}
If the integral in Eq. (\ref{ZI2a}) converges for any $N$,
by performing saddle point integration we see that the saddle point
${{m}}^{\mathrm{sp}}$ is solution of the equation
\begin{eqnarray}
\label{ZI5}
{{m}}^{\mathrm{sp}}=m_0 \left(\beta J_0^{(I)},
\beta J^{(I)}{{m}}^{\mathrm{sp}} + \beta h\right), 
\end{eqnarray}
so that, if the stability condition
\begin{eqnarray}
\label{ZI6}
1- \beta J^{(I)}
\tilde{\chi}_0 \left(\beta J_0^{(I)},\beta J^{(I)}{{m}}^{\mathrm{sp}} + \beta h\right)>0, 
\end{eqnarray}
is satisfied, in the thermodynamic limit we arrive at the following
expression for the free energy density $f_I$ of the related Ising model 
\begin{eqnarray}
\label{ZI7}
\beta f_I = \left[\frac{\beta}{2} J^{(I)}{{m}}^2
+\beta f_0 \left(\beta J_0^{(I)},\beta J^{(I)}{{m}} 
+ \beta h\right)\right]_{{{m}}={{m}}^{\mathrm{sp}}}.
\end{eqnarray}
Similarly, any correlation function $C_I$ of the related Ising model 
is given in terms of the correlation function $C_0$ of the unperturbed
model by the following relation
\begin{eqnarray}
\label{ZI7b}
C_I =  C_0\left(\beta J_0^{(I)},\beta J^{(I)}{{m}} 
+ \beta h\right)|_{{{m}}={{m}}^{\mathrm{sp}}}.
\end{eqnarray}

Of course, the saddle point solution ${{m}}^{\mathrm{sp}}$ represents the magnetization
of the related Ising model, 
as can be checked directly by deriving
Eq. (\ref{ZI7}) with respect to $\beta h$ and by using Eq. (\ref{ZI5}).

If the saddle point equation (\ref{ZI5}) has more stable solutions,
the ``true'' free energy and the ``true'' observable of the related Ising model 
will be given by Eqs. (\ref{ZI7}) and (\ref{ZI7b}), respectively, calculated 
at the saddle point solution which minimizes Eq. (\ref{ZI7}) itself
and that we will indicate with $m_I$.

Let us call $\beta_{c0}^{(I)}$ the inverse critical temperature of the unperturbed
model with coupling $J_0^{(I)}$ and zero external field, 
possibly with $\beta_{c0}^{(I)}=\infty$ if no phase transition exists.
As stressed in Sec. IIIB, for the unperturbed model 
we use the expression ``critical temperature'' for 
any temperature where the magnetization $m_0$ at
zero external field passes from 0 to a non zero value, continuously or not.
Note that, as a consequence, if $J_0^{(I)}<0$,
we have formally $\beta_{c0}^{(I)}=\infty$, 
independently from the fact that
some antiferromagnetic order may be not zero.

Let us start to make the obvious observation that a necessary condition for the related Ising model to
have a phase transition at $h=0$ and for a finite temperature, 
is the existence of some paramagnetic region P$_I$ where
$m_I=0$. We see from the saddle point equation (\ref{ZI5}) that, for $h=0$,
a necessary condition for $m_I=0$ to be a solution is that be
$\beta\leq\beta_{c0}^{(I)}$ for any $\beta$ in P$_I$, 
from which we get also $\beta_c^{(I)}\leq\beta_{c0}^{(I)}$. 
In a few lines we will see however that the inequality must be strict 
if $\beta_{c0}^{(I)}$ is finite, which in particular excludes
the case $J_0<0$ (for which the inequality to be proved is trivial).

Let us suppose for the moment that be $\beta_c^{(I)}<\beta_{c0}^{(I)}$.
For $\beta<\beta_{c0}^{(I)}$
and $h=0$, the saddle point equation (\ref{ZI5}) has always the trivial 
solution $m_I=0$ which, according to Eq. (\ref{ZI6}), is also a stable solution if
\begin{eqnarray}
\label{ZI8}
1- \beta J^{(I)}
\tilde{\chi}_0 \left(\beta J_0^{(I)},0\right)>0.
\end{eqnarray}
The solution $m_I=0$ starts to be unstable when 
\begin{eqnarray}
\label{ZI9}
1- \beta J^{(I)}
\tilde{\chi}_0 \left(\beta J_0^{(I)},0\right)=0.
\end{eqnarray}
Eq. (\ref{ZI9}), together with the constrain $\beta_c^{(I)}\leq\beta_{c0}^{(I)}$,
gives the critical temperature of the related Ising model $\beta_c^{(I)}$.
In the region of temperatures where Eq. (\ref{ZI8}) is violated, 
Eq. (\ref{ZI5}) gives two symmetrical stable 
solutions $\pm m_I\neq 0$. 
From Eq. (\ref{ZI9}) we see also that the case $\beta_c^{(I)}=\beta_{c0}^{(I)}$
is impossible unless be $J^{(I)}=0$, since the susceptibility 
$\tilde{\chi}_0(\beta J_0^{(I)},0)$ must diverge at $\beta_{c0}^{(I)}$.
We have therefore proved that $\beta_c^{(I)}<\beta_{c0}^{(I)}$.
Note that for $J_0^{(I)}\geq 0$ and $\beta<\beta_{c0}^{(I)}$ 
Eq. (\ref{ZI8}) is violated only for
$\beta>\beta_c^{(I)}$, whereas for $J_0^{(I)}<0$ Eq. (\ref{ZI8}) 
in general may be violated also in finite regions of the $\beta$ axis.

The critical behavior of the related Ising model 
can be studied by expanding Eq. (\ref{ZI5}) for small fields.
However, we find it more convenient to expand $L({{m}})$ in series around ${{m}}=0$
since in this way everything can be cast in the standard formalism of 
the Landau theory of phase transitions. From Eq. (\ref{ZI2}),
taking into account that the function $\tilde{\chi}_0 \left(\beta J_0,\beta h\right)$
is an even function of $\beta h$, we have the following general expression
valid for any $m$, $\beta$ and small $h$
\begin{eqnarray}
\label{ZI10}
L(m)= \beta f_0 \left(\beta J_0^{(I)},0\right) - m_0\left(\beta J_0^{(I)},0\right)\beta h +
\psi\left(m\right),
\end{eqnarray}
where we have introduced the Landau free energy density $\psi(m)$ given by
\begin{eqnarray}
\label{ZI11}
\psi\left(m\right)&=& \frac{1}{2}a m^2 + \frac{1}{4} b m^4+\frac{1}{6} c m^6
\nonumber \\ &&
 - m \beta \tilde{h}
+\Delta \left(\beta f_0\right)\left(\beta J_0^{(I)},\beta J^{(I)}m \right),
\end{eqnarray}
where 
\begin{eqnarray}
\label{ZI12}
a=\left[1-\beta J^{(I)}\tilde{\chi}_0 \left(\beta J_0^{(I)},0\right)\right]\beta J^{(I)},
\end{eqnarray}
\begin{eqnarray}
\label{ZI13}
b=- \frac{\partial^2}{\partial(\beta h)^2}
{\left. {\tilde{\chi}_0\left(\beta J_0^{(I)},\beta h\right)}
\right|_{_{\beta h =0} }
\frac{\left(\beta J^{(I)}\right)^4}{3!}},
\end{eqnarray}
\begin{eqnarray}
\label{ZI14}
c=- \frac{\partial^4}{\partial(\beta h)^4}
{\left. {\tilde{\chi}_0\left(\beta J_0^{(I)},\beta h\right)}
\right|_{_{\beta h =0} }
\frac{\left(\beta J^{(I)}\right)^6}{5!}},
\end{eqnarray}
\begin{eqnarray}
\label{ZI15}
\tilde{h}=m_0\left(\beta J_0^{(I)},0\right)J^{(I)}+
\tilde{\chi}_0\left(\beta J_0^{(I)},0\right)\beta ^{(I)} J^{(I)}\beta h,
\end{eqnarray}
finally, the last term $\Delta \left(\beta f_0\right) \left(\beta J_0^{(I)},\beta J^{(I)}m\right)$
is defined implicitly to render Eqs. (\ref{ZI10}) and (\ref{ZI11}) exact, 
but terms $\mathop{O}(h^2)$ and $\mathop{O}(m^3 h)$; explicitly
\begin{eqnarray}
\label{ZI16}
&&\Delta \left(\beta f_0\right) \left(\beta J_0^{(I)},\beta J^{(I)}m\right)=
\nonumber \\
&-&\sum_{k=4}^\infty \frac{\partial^{2k-2}}{\partial(\beta h)^{2k-2}}
{\left. {\tilde{\chi}_0\left(\beta J_0^{(I)},\beta h\right)}
\right|_{_{\beta h =0} }
\frac{\left(\beta J^{(I)}\right)^{2k}}{(2k)!}}.
\end{eqnarray}
 
Finally, to come back to the original random model, we have just
to perform the reversed mapping transformations (\ref{MT3}) and (\ref{MT4}) 
in Eqs. (\ref{ZI2})-(\ref{ZI16}). As a result we get 
immediately Eqs. (\ref{THEOa})-(\ref{THEOf1}), but Eq. (\ref{THEOl}).

\section{Derivation of Eq. (\ref{THEOl}) and  Eqs. (\ref{THEOrule})-(\ref{THEOrule9})}
Concerning Eq. (\ref{THEOl}) for the full expression of the free energy
density, it can be obtained by using 
Eqs. (\ref{0logZ2}), (\ref{varphi}), (\ref{mapp2g}) and (\ref{mapp3g}).
Here $\varphi_I$ is the high temperature part of the free energy
density of the related Ising model we have just solved:
\begin{eqnarray}
\label{ZI17} 
&& - \beta f_I = 
 \lim_{N\to\infty}\frac{1}{N}\sum_{(i,j)\in\Gamma_0}\log\left[\cosh(\beta J_0^{(I)})\right]
+\frac{N-1}{2}\nonumber \\
&& \times \log\left[\cosh\left(\beta J^{(I)}/N\right)\right]+
\log\left[2\cosh(\beta h)\right] + \varphi_I
\end{eqnarray}
where we have taken into account the fact that our related Ising model
has $|\Gamma_0|$ connections with coupling $J_0^{(I)}$ and 
$N(N-1)/2$ connections with the coupling $J^{(I)}/N$.
By using Eq. (\ref{ZI7}) calculated in $m_I$ and Eq. (\ref{ZI17}),
for large $N$ we get
\begin{eqnarray}
\label{ZI18}
&& \varphi_I = - \frac{\beta}{2} J^{(I)}{{m_I}}^2
- \beta f_0 \left(\beta J_0^{(I)},\beta J^{(I)}{{m_I}} +\beta h\right)\nonumber \\
&& - \lim_{N\to\infty}\frac{1}{N}\sum_{(i,j)\in\Gamma_0}
\log\left[\cosh(\beta J_0^{(I)})\right] \nonumber \\
&& -\log\left[2\cosh(\beta h)\right] + \mathop{O}\left(\frac{1}{N}\right).
\end{eqnarray}
Therefore, on using Eq. (\ref{mapp2g}), for the non trivial part
$\varphi^{(\Sigma)}$ of the random system, 
up to corrections $\mathop{O}\left(1/N\right)$, we arrive at 
\begin{eqnarray}
\label{ZI19}
&& \varphi^{(\Sigma)} = - \frac{\beta}{2}
J^{(\Sigma)}\left(m^{(\Sigma)}\right)^2
-\log\left[2\cosh(\beta h)\right]
\nonumber \\
&& - \lim_{N\to\infty}\frac{1}{N}\sum_{(i,j)\in\Gamma_0}
\log\left[\cosh(\beta J_0^{(\Sigma)})\right]
\nonumber \\
&& - \beta f_0 \left(\beta J_0^{(\Sigma)},\beta J^{(\Sigma)}m^{(\Sigma)} +\beta h\right) 
\end{eqnarray}
In terms of the function $L^{(\Sigma)}(m)$ Eq. (\ref{ZI19}) reads as
\begin{eqnarray}
\label{ZI20}
&& \varphi^{(\Sigma)} = - L^{(\Sigma)}\left(m^{(\Sigma)}\right)
\nonumber \\
&& - \lim_{N\to\infty}\frac{1}{N}\sum_{(i,j)\in\Gamma_0}
\log\left[\cosh(\beta J_0^{(\Sigma)})\right]
\nonumber \\
&&
-\log\left[2\cosh(\beta h)\right].
\end{eqnarray}
By using Eqs. (\ref{0logZ2}), (\ref{ZI20}), (\ref{mapp2g}) and (\ref{mapp3g}),
with $l=1$ or $2$ for $\Sigma$=F or $\Sigma$=SG, respectively,
we get Eq. (\ref{THEOl}).
 
For $h=0$ Eq. (\ref{ZI20}) can conveniently be rewritten also as
\begin{eqnarray}
\label{ZI21}
&& \varphi^{(\Sigma)} = \varphi_0\left(\beta J_0^{(\Sigma)},0\right)
\nonumber \\
&& + \left[L^{(\Sigma)}\left(0\right)
-L^{(\Sigma)}\left(m^{(\Sigma)}\right)\right],
\end{eqnarray}
where 
\begin{eqnarray}
&& \varphi_0(\beta J_0,\beta h)=
-\beta f_0 \left(\beta J_0^{(\Sigma)},\beta h\right)
\nonumber \\
&& - \lim_{N\to\infty}\frac{1}{N}\sum_{(i,j)\in\Gamma_0}
\log\left[\cosh(\beta J_0^{(\Sigma)})\right]
\nonumber \\
&& -\log\left[2\cosh(\beta h)\right],
\end{eqnarray}
is the high temperature part of the free energy
density of the unperturbed model with coupling $J_0^{(\Sigma)}$ and external field $h$. 
There are some important properties for the function $\varphi_0(\beta J_0,0)$:
it is a monotonic increasing function of $\beta J_0$;
if the lattice $\mathcal{L}_0$ has only loops of even length, 
$\varphi_0(\beta J_0,0)$ is an even function of $\beta J_0$;
furthermore,
if $d_0<2$ and the coupling-range is finite, 
or if $d_0=\infty$ at least in a wide sense
\cite{MOIII}, in the thermodynamic limit we have $\varphi_0(\beta J_0,0)=0$;
if instead $2\leq d_0<\infty$, $\varphi_0(\beta J_0,0)\neq 0$. 
We see here therefore what anticipated in Sec. IVA: when $J_0\neq 0$,
the symmetry among the random couplings is broken and for $d_0$
sufficiently high this reflects in a non zero $\varphi^{(\Sigma)}$ 
also in the P region.

Next we prove Eqs. (\ref{THEOrule})-(\ref{THEOrule9}).
To this aim we have to calculate Eq. (\ref{ZI21}) at the
leading solution $\bar{m}^{(\Sigma)}$ and to compare $\varphi^{\mathrm{(F)}}$ and $\varphi^{\mathrm{(SG)}}$.
Note that the term in the square parenthesis of Eq. (\ref{ZI21})
is non negative since $\bar{m}^{(\Sigma)}$ is the absolute minimum of $L^{(\Sigma)}$.
We recall that for critical temperature we mean here any temperature
lying on the boundary P-F or P-SG, 
so that $\bar{m}^{(\Sigma)}|_\beta=0$ for any $\beta$ in the P region.

\subsection{$J_0\geq 0$}
If $J_0\geq 0$, for both the solution with label F and SG, 
we have only one second order phase transition so that
$\bar{m}^{\mathrm{(F)}}=0$ and $\bar{m}^{\mathrm{(SG)}}=0$, respectively, are 
the stable and leading solutions even on the boundary with the P region.

Let us suppose $\beta_c^{\mathrm{(F)}}<\beta_c^{\mathrm{(SG)}}$.
Let be $\varphi_0(\cdot,0)\neq 0$.
From Eq. (\ref{ZI21}) and by using 
$J_0^{\mathrm{(F)}}>J_0^{\mathrm{(SG)}}$, we see that
\begin{eqnarray}
\label{ZI22}
&& \varphi^{(\mathrm{F})}|_{\beta_c^{\mathrm{(F)}}} 
= \varphi_0\left(\beta_c^{\mathrm{(F)}} J_0^{\mathrm{(F)}},0\right)
\nonumber \\
&&> \varphi^{(\mathrm{SG})}|_{\beta_c^{\mathrm{(F)}}} 
= \varphi_0\left(\beta_c^{\mathrm{(F)}} J_0^{\mathrm{(SG)}},0\right).
\end{eqnarray}
Finally, by using this result and the general rule given by Eqs. (\ref{mapp2g}) and
(\ref{mapp3g}), we see 
(and with a stronger reason, due to the factor 1/2 appearing in these equations for the SG solution) that 
the stable phase transition is the P-F one:
$\beta_c=\beta_c^{\mathrm{(F)}}$. 
Similarly, by using Eq. (\ref{ZI21})
for $\beta_c^{\mathrm{(F)}}<\beta<\beta_c^{\mathrm{(SG)}}$, 
we see that even for any $\beta$ in the interval 
$(\beta_c^{\mathrm{(F)}},\beta_c^{\mathrm{(SG)}})$
the stable solution is that with label F. This last observation makes also clear that if
$\varphi_0(\cdot,0)=0$ we reach the same conclusion: F is the stable
phase in all the region
$\beta_c^{\mathrm{(F)}}<\beta<\beta_c^{\mathrm{(SG)}}$ and in particular this 
implies also that the stable phase transition is the P-F one: 
$\beta_c=\beta_c^{\mathrm{(F)}}$.

Let us suppose $\beta_c^{\mathrm{(F)}}>\beta_c^{\mathrm{(SG)}}$.
If $\varphi_0(\cdot,0)\neq 0$, we arrive at 
\begin{eqnarray}
\label{ZI23}
&& \varphi^{(\mathrm{F})}|_{\beta_c^{\mathrm{(SG)}}} 
= \varphi_0\left(\beta_c^{\mathrm{(SG)}} J_0^{\mathrm{(F)}},0\right)
\nonumber \\
&&> \varphi^{(\mathrm{SG})}|_{\beta_c^{\mathrm{(SG)}}} 
= \varphi_0\left(\beta_c^{\mathrm{(SG)}} J_0^{\mathrm{(SG)}},0\right).
\end{eqnarray}
Finally, by using this result and the general rule given by Eqs. (\ref{mapp2g}) and
(\ref{mapp3g}), we see (and with a stronger reason) that 
the stable phase on the boundary is that predicted by the F solution which
has zero magnetization at $\beta_c^{\mathrm{(SG)}}$. This does not
imply that $\beta_c=\beta_c^{\mathrm{(F)}}$, but only that
$\beta_c^{\mathrm{(SG)}}<\beta_c\leq \beta_c^{\mathrm{(F)}}$.
If instead $\varphi_0(\cdot,0)=0$,
by using Eq. (\ref{ZI21})
for $\beta_c^{\mathrm{(SG)}}<\beta<\beta_c^{\mathrm{(F)}}$, 
we see that for any $\beta$ in the interval 
$(\beta_c^{\mathrm{(SG)}},\beta_c^{\mathrm{(F)}})$
the stable solution is SG and then, in particular, the stable boundary is P-SG:
$\beta_c=\beta_c^{\mathrm{(SG)}}$.

\subsection{$J_0<0$}
If $J_0< 0$, for the solution with label F, we may have both first and second order phase transitions.
In the first case we cannot in general assume that 0 is the stable and leading
solution on the boundary with the P region:
$m^\mathrm{(F)}|_{\beta_c^\mathrm{(F)}}\neq 0$ in general.
As a consequence, for a first order transition the term in square
parenthesis of Eq. (\ref{ZI21}) may be non zero even on the critical surface. 
Furthermore, as $J_0<0$, for the solution F we have at least two critical temperatures 
that we order as $\beta_{c1}^\mathrm{(F)}\geq \beta_{c2}^\mathrm{(F)}$.
However, despite of these complications, 
if we assume that $\mathcal{L}_0$ has only loops of even length,
$\varphi_0(\cdot,0)$ turns out to be an even function and,
due to the inequality $|J_0^{\mathrm{(F)}}|>J_0^{\mathrm{(SG)}}$,
almost nothing changes in the arguments we have used in the previous case $J_0\geq 0$.

Let us consider first the surfaces $\beta_{c2}^{\mathrm{(F)}}$
and $\beta_{c}^{\mathrm{(SG)}}$. Independently of the kind of phase
transition, first or second order, we arrive again at Eqs. (\ref{ZI22})
and (\ref{ZI23}), for $\beta_{c2}^{\mathrm{(F)}}<\beta_{c}^{\mathrm{(SG)}}$ 
and $\beta_{c2}^{\mathrm{(F)}}>\beta_{c}^{\mathrm{(SG)}}$, respectively,
with the same prescription for the cases $\varphi_0(\cdot,0)\neq 0$,
or $\varphi_0(\cdot,0)= 0$. 

Let us now consider the surfaces $\beta_{c1}^{\mathrm{(F)}}$
and $\beta_{c}^{\mathrm{(SG)}}$. If $\beta_{c1}^{\mathrm{(F)}}<\beta_{c}^{\mathrm{(SG)}}$
and $\varphi_0(\cdot,0)\neq 0$,
for any $\beta$ in the interval $[\beta_{c1}^{\mathrm{(F)}},\beta_{c}^{\mathrm{(SG)}}]$ we have, 
\begin{eqnarray}
\label{ZI24}
&& \varphi^{(\mathrm{F})}|_{\beta} = \varphi_0\left(\beta J_0^{\mathrm{(F)}},0\right)
\nonumber \\
&&> \varphi^{(\mathrm{SG})}|_{\beta} 
= \varphi_0\left(\beta J_0^{\mathrm{(SG)}},0\right),
\end{eqnarray}
so that the interval $[\beta_{c1}^{\mathrm{(F)}},\beta_{c}^{\mathrm{(SG)}}]$
is a stable P region corresponding to the solution with label F.
Similarly, we arrive at the same conclusion if $\varphi_0(\cdot,0)=0$.
However, the interval of temperatures where the P region 
$[\beta_{c1}^{\mathrm{(F)}},\beta_{c}^{\mathrm{(SG)}}]$ is stable
can be larger when $d_0\geq 2$. In fact (exactly as we have seen for $J_0\geq 0$) 
in this case the P-SG stable boundary may stay at lower temperatures.
Finally, let us consider the case $\beta_{c1}^{\mathrm{(F)}}>\beta_{c}^{\mathrm{(SG)}}$.
If $\varphi_0(\cdot,0)= 0$, by using 
Eq. (\ref{ZI21}) we see that for any $\beta> \beta_{c}^{\mathrm{(SG)}}$
we have that the stable solution corresponds to the SG one, 
so that there is no stable boundary with the P region. 
If instead $\varphi_0(\cdot,0)\neq 0$, 
due to the fact that $\varphi^{(\mathrm{SG})}|_{\beta}$ 
and $\varphi^{(\mathrm{F})}|_{\beta}$ grow in a different way with $\beta$,
we are not able to make an exact comparison, and it is possible that 
the P-F boundary becomes stable starting from some $\beta_{c1}$ with 
$\beta_{c1}\geq \beta_{c1}^{\mathrm{(F)}}$. In general, as in the case $J_0<0$, we could have 
one (or even more) sectors where the P region corresponding to the solution
with label F is stable. 

\section{Conclusions}
In this paper we have presented a novel and general method
to face analytically random Ising models defined over small-world networks.
The key point of our method lies on the fact that, at least in the P region,
any such a model can be exactly mapped on a suitable fully connected model,
whose resolvability is in general non trivial for $d_0>1$, but still as feasible
as a non random model. The main result stands then in deriving a general
self-consistent equation, Eq. (\ref{THEOa}), able to describe effectively the model 
and based on the knowledge of $m_0(\beta J_0,\beta h)$, the magnetization
of the unperturbed model in the presence of an arbitrary external field.

The physical interpretation of the these results is straightforward:
from Eqs. (\ref{THEOa}) we see that, concerning the magnetization $m^{(\mathrm{F})}$,
the effect of adding long range Poisson distributed bonds
implies that the system - now perturbed - besides the coupling $J_0$, 
feels also an effective external field
$J^{(\mathrm{F})}$ shrunk
by $m^{(\mathrm{F})}$ itself; whereas, concerning $m^{(\mathrm{SG})}$, 
the effect is that the system
now feels a slightly modified 
effective coupling $\beta J_0^{(\mathrm{SG})}$ and an effective 
external field 
$J^{(\mathrm{SG})}$ shrunk
by $m^{(\mathrm{SG})}$ itself. 

We are therefore in the presence of an
\textit{effective field theory} which, as opposed to a simpler 
\textit{mean-field theory},
describes $m^{(\mathrm{F})}$ and $m^{(\mathrm{SG})}$ in terms of, not only
an effective external field, but also in terms of the non trivial function 
$m_0(\beta J_0,\beta h)$ which, in turn,
takes into account the correlations
due to the non zero short-range coupling $J_0$ or $J_0^{(\mathrm{SG})}$ felt
by the unperturbed model. The combination of these two effects gives rise to
the typical behavior of models defined over small-world networks:
the presence of a non zero effective external field causes the existence
of a mean-field phase transition also in low $d_0$ dimension. 
However, the precise determination
of both the critical surface and the correlation functions is determined in a non
trivial way by the unperturbed magnetization $m_0(\beta J_0,\beta h)$. 

In this paper we have then applied the method to analyze, in the most general
feasible situation, the critical behavior of generic models with $J_0\geq 0$
and $J_0<0$, showing that they give rise to two strictly different scenario of phase
transitions. In the first case, we have a mean-field second-order phase transition
but with a finite correlation length. Whereas, in the second case, 
we have multiple first and second order phase transitions. 
Furthermore, we have shown that the combination of the F and SG solution,
results in a total of four possible kind of phase diagrams according to the cases
\textbf{(1)} $(J_0\geq 0;~d_0<2,~\mathrm{or}~d_0=\infty)$, \textbf{(2)} $(J_0\geq 0;~2\leq d_0<\infty)$,
\textbf{(3)} $(J_0< 0;~d_0<2,~\mathrm{or}~d_0=\infty)$, and \textbf{(4)} $(J_0< 0;~2\leq d_0<\infty)$.  
One remarkable difference between the cases with $d_0<2,~\mathrm{or}~d_0=\infty$ and the cases with
$2<d_0<\infty$, is that in the latter we have also first-order P-SG transitions
and, moreover, re-entrance phenomena are in principle possible even for $J_0\geq 0$.

In a forthcoming paper (part II) we will apply the method to study 
in detail several solvable (by our method) models of interest. 

As discussed in Sec. IIIE, the method can be readily generalized to study
more complex phase diagrams that may emerge when one considers
couplings $J_{0}$ depending on the bond $b$, and to study 
small-world antiferromagnetism.

Models defined on complex small-world networks \cite{Review} are 
interesting subject of future works.


\begin{acknowledgments}
This work was supported by the FCT (Portugal) grants
SFRH/BPD/24214/2005, pocTI/FAT/46241/2002 and
pocTI/FAT/46176/2003, and the Dysonet Project.
We thank A. L. Ferreira and A. Goltsev for useful discussions.
\end{acknowledgments}

\appendix
\section{Generalization to non homogeneous external field}
In this appendix we prove Eq. (\ref{THEOC2b}) calculating 
the $\mathop{O}(1/N)$ correction responsible for
the divergence of the susceptibility of the random system at $T_c$.
To this aim we firstly need to generalize our method to
an arbitrary external field. 
Let us consider again a fully connected model having - as done in Sec. V -
long-range couplings $J$ (for brevity we will here omit the label $I$)
and short-range couplings $J_0$ but now
immersed in an arbitrary (non homogeneous) external field $\{h_n\}$, 
where $n=1,\ldots,N$. 
After using the Gaussian transformation 
we have the following partition function: 
\begin{eqnarray}
\label{App1}
Z&=& c_N \int_{-\infty}^{\infty} d{{m}} ~ e^{-N L({{m}})},
\end{eqnarray}
where we have introduced the function 
\begin{eqnarray}
\label{App2}
L({{m}})= \frac{\beta}{2} J{{m}}^2 
+\beta f_0 \left(\beta J_0,\{\beta J{{m}} + \beta h_n\}\right),
\end{eqnarray}
$f_0 \left(\beta J_0,\{\beta h_n\}\right)$
being the free energy density of the unperturbed model
in the presence of an arbitrary external field $\{\beta h_n\}$. 
By using 
\begin{eqnarray}
\label{App3}
\partial_{\beta h_i}~\beta f_0(\beta J_0,\{\beta h_n\})=
-m_{0i}(\beta J_0,\{\beta h_n\}), 
\end{eqnarray}
and
\begin{eqnarray}
\label{App4}
\partial_{\beta h_j}~m_{0i}(\beta J_0,\{\beta h_n\})&\equiv&
\tilde{\chi}_{0;i,j}(\beta J_0,\{\beta h_n\})\nonumber \\ 
&=&\media{\sigma_i\sigma_j}_0-\media{\sigma_i}_0\media{\sigma_j}_0
\end{eqnarray}
we get
\begin{eqnarray}
\label{App5}
L'({{m}})= \beta J
\left[{{m}}-\frac{1}{N}\sum_i 
m_{0i}\left(\beta J_0,\{\beta J{{m}} + \beta h_n\}\right)\right], 
\end{eqnarray}
\begin{eqnarray}
\label{App6}
L''({{m}})&=& \beta J 
\left[1- \beta J \frac{1}{N} \right. \nonumber \\ 
&\times& \left. \sum_{i,j} 
\tilde{\chi}_{0ij} \left(\beta J_0,\{\beta J{{m}} + \beta h_n\}\right)\right].
\end{eqnarray}
By performing the saddle point integration we see that the saddle point
${{m}}^{\mathrm{sp}}$ is solution of the equation
\begin{eqnarray}
\label{App7}
{{m}}^{\mathrm{sp}}=\frac{1}{N}\sum_i 
m_{0i}\left(\beta J_0,\{\beta J{{m}}^{\mathrm{sp}} + \beta h_n\}\right)
\end{eqnarray}
hence, by using 
\begin{eqnarray}
\label{App7a}
&& \tilde{\chi}_{0} \left(\beta J_0,\{\beta J{{m}} + \beta h_n\}\right) = 
\nonumber \\ &&\frac{1}{N}  
\sum_{i,j}\tilde{\chi}_{0ij} \left(\beta J_0,\{\beta J{{m}} + \beta h_n\}\right),
\end{eqnarray}
we see that if the stability condition
\begin{eqnarray}
\label{App8}
1- \beta J
\tilde{\chi}_0 \left(\beta J_0,\{\beta J{{m}}^{\mathrm{sp}} + \beta h_n\}\right)>0, 
\end{eqnarray}
is satisfied, in the thermodynamic limit we arrive at the following
expression for the free energy density $f$ of the related Ising model 
immersed in an arbitrary external field
\begin{eqnarray}
\label{App9}
\beta f = \left[\frac{\beta}{2} J{{m}}^2
+\beta f_0 \left(\beta J_0,\{\beta J{{m}}+ \beta h_n\}\right)\right]_{{{m}}={{m}}^{\mathrm{sp}}}.
\end{eqnarray}
On the other hand, by derivation with respect to $\beta h_i$ and 
by using Eq. (\ref{App7}), it is immediate to verify that
\begin{eqnarray}
\label{App10}
m_i\equiv \media{\sigma_i}=m_{0i}(\beta J_0,\{\beta J{{m}}^{\mathrm{sp}} + \beta h_n\}),
\end{eqnarray}
and then also (from now on for brevity on we omit the symbol $^{\mathrm{sp}}$)
\begin{eqnarray}
\label{App11}
m=\frac{1}{N}\sum_i m_i.
\end{eqnarray}

We want now to calculate the correlation functions.
From Eq. (\ref{App10}), by deriving with respect to $\beta h_j$ we have
\begin{eqnarray}
\label{App12}
\tilde{\chi}_{ij} &\equiv& \frac{\partial m_i}{\partial (\beta h_j)}=
\sum_l \tilde{\chi}_{0;i,l} (\beta J_0,\{\beta J{{m}}+ \beta h_n\})
\nonumber \\ 
&\times & \left(\beta J\frac{\partial m}{\partial (\beta h_j)} + \delta_{l,j}\right),
\end{eqnarray}
which, by summing over the index $i$ and using (\ref{App11}), gives
\begin{eqnarray}
\label{App13}
\frac{\partial m}{\partial (\beta h_j)}=
\frac{\frac{1}{N}\sum_i \tilde{\chi}_{0;i,j} (\beta J_0,\{\beta J{{m}}+ \beta h_n\})}
{1-\beta J \tilde{\chi}_{0} (\beta J_0,\{\beta J{{m}}+ \beta h_n\})}.
\end{eqnarray}
We can now insert Eq. (\ref{App13}) in the rhs of Eq. (\ref{App12}) to get
\begin{eqnarray}
\label{App14}
\tilde{\chi}_{ij}&=&
\tilde{\chi}_{0;i,j} (\beta J_0,\{\beta J{{m}}+ \beta h_n\}) \nonumber \\
&+& \frac{\beta J}{N}\frac{
\sum_l \tilde{\chi}_{0;l,j} 
\sum_k \tilde{\chi}_{0;i,k}} 
{1-\beta J \tilde{\chi}_{0} (\beta J_0,\{\beta J{{m}}+ \beta h_n\})},
\end{eqnarray}
where for brevity we have omitted the argument in $\tilde{\chi}_{0;l,j}$ 
and $\tilde{\chi}_{0;i,k}$, which is the same of $\tilde{\chi}_{0}$ appearing
in the denominator.
If we now come back to choice a uniform external field $h_n=h$, $n=1,\ldots,N$, 
we can use translational invariance and for the related Ising model (fully connected) 
we obtain the following correlation function
\begin{eqnarray}
\label{App15}
\tilde{\chi}_{ij}&=&\frac{\beta J}{N}
\frac{\left[\tilde{\chi}_{0} (\beta J_0,\beta J{{m}}+ \beta h)\right]^2}
{1-\beta J \tilde{\chi}_{0} (\beta J_0,\beta J{{m}}+ \beta h)}
\nonumber \\
&+& \tilde{\chi}_{0;i,j} (\beta J_0,\beta J{{m}}+ \beta h).
\end{eqnarray}
Finally, by performing the mapping substitutions (\ref{MT3}) and (\ref{MT4})
we arrive at Eq. (\ref{THEOC2b}). 

Similarly, any correlation function $C$ of the related Ising model 
will be given by a similar formula with the leading term $C_0$ plus
a correction $\mathop{O}(1/N)$ becoming important only near $T_c$.


\end{document}